\documentclass[preprintnumbers,nofootinbib,amsmath,amssymb]{revtex4}

\def\ga{\mathrel{\raise.3ex\hbox{$>$\kern-.75em\lower1ex\hbox{$\sim$}}}}
\def\la{\mathrel{\raise.3ex\hbox{$<$\kern-.75em\lower1ex\hbox{$\sim$}}}}

\newcommand\beq{\begin{equation}}
\newcommand\eeq{\end{equation}}
\newcommand\beqar{\begin{eqnarray}}
\newcommand\eeqar{\end{eqnarray}}
\usepackage{graphicx}
\usepackage{epsfig}
\usepackage{epsf}
\usepackage{rotating}
\usepackage{verbatim}

\usepackage{bbm}

\usepackage{multirow}

\def\identity{\mathbbm{1}}

\begin{document}

\preprint{ArXiv:1001.4088} \preprint{UMN--TH--2835/10} \vskip 0.2in
\title{
Scalar-Scalar, Scalar-Tensor, and Tensor-Tensor \\

Correlators from Anisotropic Inflation
}

\author{A. Emir G\"umr\"uk\c{c}\"uo\u{g}lu, Burak Himmetoglu and Marco Peloso}
\address{{\it School of Physics and Astronomy, University of Minnesota, Minneapolis, MN 55455, USA}}

\begin{abstract}
We compute the phenomenological signatures of a model (Watanabe et al' 09) of anisotropic inflation driven by a scalar and a vector field. The action for the vector is U(1) invariant, and the model is free of ghost instabilities. A suitable coupling of the scalar to the kinetic term of the vector allows for a slow roll evolution of the vector vev, and hence for a prolonged anisotropic expansion; this provides a counter example to the cosmic no hair conjecture. We compute the nonvanishing two point correlation functions between physical modes of the system, and express them in terms of power spectra with angular dependence. The anisotropy parameter $g_*$ for the scalar-scalar spectrum (defined as in the Ackerman et al '07 parametrization) turns out to be negative in the simplest realization of the model, which, therefore, cannot account for the angular dependence emerged in some analyses of the WMAP data. A $g_*$ of order -0.1 is achieved when the energy of the vector is about $6-7$ orders of magnitude smaller than that of the scalar during inflation. For such values of the parameters, the scalar-tensor correlation (which is in principle a distinctive signature of anisotropic spaces) is smaller than the tensor-tensor correlation.
\end{abstract}

\maketitle

\section{Introduction}

The WMAP data on the CMB anisotropies \cite{Hinshaw:2008kr}
have strongly improved our knowledge of the Universe, and are in overall excellent agreement with
the theory of inflation. However, the unprecedented quality of the data has driven a number of analysis on finer effects, which seem to be hard to reconcile with the simplest inflationary models. 
These so called
`anomalies' include the low power in the quadrupole
moment~\cite{cobe,wmap1,lowl}, the alignment of the lowest
multipoles~\cite{axis}, a $\sim 5^\circ$ cold spot with suppressed
power \cite{cold}, an asymmetry in power between the northern and
southern ecliptic hemispheres~\cite{asym}, and broken rotational
invariance~\cite{Groeneboom:2008fz}. Most of these effects (and, particularly, those appearing at the largest scales) can be generally thought as violation of statistical isotropy of the CMB. It has been suggested in \cite{Gumrukcuoglu:2006xj}
that a period of anisotropic expansion during inflation may explain some of these features. 
Specifically, ref. \cite{Gumrukcuoglu:2006xj} pointed out that the anisotropic expansion would provide a primordial power spectrum with a dependence on the directionality of the modes, and computed the 
corresponding correlation matrix $\langle a_{\ell m} \, a_{\ell'm'}^* \rangle \propto \!\!\!\!\! \not \;\; \delta_{\ell \ell'} \, \delta_{m m'}$. 

Ref. \cite{Ackerman:2007nb} introduced the power spectrum parametrization
\begin{equation}
P \left( {\bf k} \right) = P \left( k \right) \left[ 1 + g_* \left( k \right) \, \xi^2 \right] 
\label{acw}
\end{equation}
where $\xi$ is the cosine of the angle between ${\bf k}$ and a given fixed direction, and $k = \vert {\bf k} \vert$, and computed the corresponding CMB correlation matrix as an expansion series in $g_*$. Although the specific inflationary model proposed in \cite{Ackerman:2007nb} to obtain this power spectrum was later found to be unstable \cite{Himmetoglu:2008zp,Himmetoglu:2008hx}, the ACW study based on
(\ref{acw}) is extremely useful, since it provides a general reference result for generic breaking of rotational invariance during inflation.~\footnote{Ref. \cite{Carroll:2008br} studied instead the power spectrum for violation of translational invariance. A general framework for directional dependence of the power spectrum was studied  in \cite{Pullen:2007tu}, where it was forecast that a quadrupole modulation of the power spectrum as small as $2\%$ of the total anisotropic signal 
can be detected by the Planck satellite.} The ACW parametrization was tested against the WMAP data in \cite{Groeneboom:2008fz}, where a $3.8 \sigma$ evidence was obtained for nonvanishing $g_*$ (an upper limit on the anisotropy was instead obtained in \cite{ArmendarizPicon:2008yr}). The study assumed a constant $g_*$, and can be likely be applied to slow roll inflationary backgrounds, in which a mild scale dependence may be expected. This analysis was then refined in two more recent works \cite{Hanson:2009gu,Groeneboom:2009cb}, which include a $\left( - i \right)^{\ell - \ell'}$ factor in the covariance matrix which was neglected in the first version of \cite{Ackerman:2007nb}, and in the analysis of \cite{Groeneboom:2008fz}. The inclusion of this factor has two important effects. Firstly, it substantially increased the significance of a nonvanishing asymmetry: $g_* = 0.29 \pm 0.031$ \cite{Groeneboom:2009cb} (the analysis includes multipoles up to $\ell = 400$ in the W-band; a smaller evidence emerges from the $V$ and $Q$ bands). Secondly, it shifted the privileged direction very close to the ecliptic poles \cite{Hanson:2009gu,Groeneboom:2009cb}. Although this near coincidence suggests that the asymmetry may not be cosmological, a systematical cause for it has not yet been determined. Specifically, ref. \cite{Groeneboom:2009cb} ruled out that the effect may be due to asymmetric beams, misestimated noise, or Zodiacal lights. 

In absence of a definite answer, it will be important to see whether the asymmetry will be also present in the forthcoming Planck data. In the meantime, it is also interesting to study whether there exist simple inflationary models which can (i) account for the asymmetry, and, possibly, (ii) provide other measurable predictions. As a first step towards this, ref. \cite{Gumrukcuoglu:2007bx} provided the formalism for computing cosmological perturbations on backgrounds with a privileged direction (Bianchi-I background, with a residual $2$d symmetry)~\footnote{Ref. \cite{Pereira:2007yy} provided the computation for general  Bianchi-I backgrounds; the results of \cite{Gumrukcuoglu:2007bx} and \cite{Pereira:2007yy} agree in the limit of $2$d isotropy, and reduce to the standard computation \cite{Mukhanov:1990me} in the limit of $3$d isotropy.}. The formalism was then applied to the simplest case 
of inflation driven by a slow rolling scalar field, assuming that the expansion rate of one direction is different from that of the other two as an initial condition at the onset of inflation. 
One of the gravity waves polarization experiences a
large growth during the initial anisotropic era (this is intimately related to
the instability of Kasner spaces), which may result in a
large $B$ signal in the CMB~\cite{Gumrukcuoglu:2008gi}.
Inflation however rapidly removes the background
anisotropy.  The modes that leave the horizon well after the universe
has isotropized were deep inside the horizon while the universe was
anisotropic, and one recovers a standard power spectrum at the
corresponding scales. The signature of the earlier anisotropic stage
are therefore present only on relatively large scales, which were
comparable to the horizon when the universe was anisotropic. 
These scales can be visible today only if the duration of inflation is limited 
to a minimal amount.

To avoid this tuning, one can obtain prolonged anisotropic inflationary
solutions by introducing some ingredients that violate the premises of
Wald's theorem~\cite{wald} on the rapid isotropization of Bianchi
universes. This has been realized through the addition of quadratic
curvature invariants to the gravity action~\cite{barrow}, with the use
of the Kalb-Ramond axion~\cite{nemanja} \footnote{For later studies of 
inflationary models with $p-$forms, see \cite{pforms}.}, or of vector
fields~\cite{ford}. A number  of  recent works focuses on this last possibility,
which is probably technically simpler than the other two.~\footnote{Several
recent works \cite{nongauss} also discuss the possible non-gaussian signature from
vector fields during inflation (with or without anisotropic expansion), as this can
be a way to differentiate them from scalar fields. See also \cite{Dimopoulos:2006ms}
for an earlier study of the cosmological curvature perturbations generated by a vector field.} The underlying idea
is that the vector has a nonvanishing expectation value along one
spatial direction, causing that direction to expand differently from the other two.
\footnote{While we focus on primordial inflation, vector fields with nonvanishing spatial vev have also been employed as sources of the late time time acceleration~\cite{dark}.}
In the standard case, i.e. for a minimal kinetic term ($-F^2/4$) and no potential term for the vector,
the vector vev rapidly decreases with the expansion of the universe, leading to a rapid
isotropization. Therefore, the action for the vector needs to be carefully arranged. To
our knowledge, four different possibilities have been explored. 
The first three possibilities are characterized by (i) a potential $V \left( A^2
\right)$ for the vector \cite{ford}, (ii) a fixed spatial norm of the
vector, enforced by a lagrange multiplier \cite{Ackerman:2007nb}, or (iii) a
nonminimal coupling of the vector to the scalar curvature \cite{Golovnev:2008cf}. \footnote{Vectors with nonminimal coupling to the curvature on an isotropic inflationary background
were studied in~\cite{AR}.} All these proposals break the U(1) symmetry which is present for a minimal vector action.
This introduces an additional degree of freedom (the longitudinal vector mode), which, 
for all of these models, turns out to be 
a ghost \cite{Himmetoglu:2008zp,Himmetoglu:2008hx,Himmetoglu:2009qi}. This leads to instabilities of
these models both at the linear (the linearized solutions for the perturbations diverge at a finite time close to horizon crossing) and nonlinear level (vacuum decay, with a UV divergent rate, into ghost-nonghost excitations).~\footnote{Refs. \cite{Dimopoulos:2008yv,Golovnev:2009rm} argued against this instability for models of the type (iii). We however believe that the arguments and the calculations presented in \cite{Himmetoglu:2008zp,Himmetoglu:2008hx,Himmetoglu:2009qi} are robust. For other stability studies, see~\cite{stability}.}

A completely different model was proposed in \cite{Watanabe:2009ct}. It is characterized by a scalar inflaton field, with a flat potential, and by a vector whose kinetic term is multiplied by a function of the scalar, $-f \left( \phi \right)^2 F^2 / 4 \,$. \footnote{Ref. \cite{Kanno:2009ei} studied the generation of primordial magnetic fields in the case in which the vector is the electromagnetic potential. An analogous study was performed in \cite{Demozzi:2009fu}.} As shown in \cite{Watanabe:2009ct}, for a suitable choice of $f$ the vev of the vector evolves slowly during inflation, and the model therefore supports a prolonged anisotropic stage. \footnote{A similar idea is also proposed in \cite{Dimopoulos:2009am}, where the function multiplying the kinetic term is taken to be an external function of time. For other works with vector fields with nonstandard kinetic terms during inflation, see \cite{nonst-kin}.} Moreover, since the model is U(1) invariant, the problematic longitudinal vector is absent. One therefore expects this model to be stable. This was indeed shown
to be the case in ref. \cite{Himmetoglu:2009mk}.~\footnote{Ref. \cite{Himmetoglu:2009mk} also presented the power spectrum of a subset of the perturbations of the model, including one of the gravity waves polarization, for which the computation is technically simpler than the complete one given here.}
We believe that this makes the model of \cite{Watanabe:2009ct} particularly interesting, since it is a complete counter example to the Wald's no hair theorem~\cite{wald}, which is proven to be free of instabilities. In the present work, we study the phenomenology of this model. We compute the full spectrum of perturbations, and  the resulting two point correlation functions. We particularly focus on two phenomenological signatures: the angular dependence of the scalar-scalar correlator, to see whether it can reproduce the value for $g_*$ reported in \cite{Groeneboom:2009cb}, and the scalar-tensor correlator, which is a distinctive prediction for this class of models (since it vanishes in the standard case). While, to our knowledge, the one  obtained here is the first definite prediction for  an anisotropic inflationary evolution, we show that the specific model proposed in \cite{Watanabe:2009ct} does not reproduce the observed signal.

The plan of this paper is the following. In Section \ref{Sec: background} we review the model of 
\cite{Watanabe:2009ct} and we study the background evolution. In Section \ref{Sec: background}
we instead perform the computation of the perturbations and we show the resulting power spectra. The two main steps are the computation of the quadratic action for the perturbations (from which we obtain the linearized evolution equations for the modes), and the quantization of that action (which is needed for computing the initial conditions for the modes, and the  correlators). For the latter, we employ the formalism of \cite{Nilles:2001fg} for the quantization of a system of coupled bosonic fields. Our results are discussed in the concluding Section \ref{sec:conclusion}.

\section{The Model and the Background Evolution}

\label{Sec: background}

We study the background solution in the model of \cite{Watanabe:2009ct}, which  is characterized by the action
\begin{equation}
S = \int d^3 x \sqrt{-g} \left[ \frac{M_p^2}{2} R - \frac{1}{2} \left( \partial \phi \right)^2 - V \left( \phi \right) - \frac{1}{4} \, f \left( \phi \right)^2 F^2 \right]
\label{action}
\end{equation}
Namely, there is a scalar field $\phi$ with potential $V\left( \phi \right)$, which is taken sufficiently flat to allow for a slow evolution of the scalar. There is also a vector field, which enters only with its kinetic term ($F_{\mu \nu} = \partial_\mu A_\nu - \partial_\nu A_\mu$), so that the action is U(1) invariant. This kinetic term is multiplied by a function of the scalar, which causes a nonstandard evolution for the vector vev. For an appropriate choice of $f \left( \phi \right)$ the energy density of the vector also evolves slowly, and the model admits a prolonged anisotropic inflationary solution \cite{Watanabe:2009ct}.

We assume a Bianchi-I background with a residual isotropy in the plane perpendicular to the vector vev,
\begin{equation}
d s^2 = - d t^2 + a \left( t \right)^2 d x^2 + b \left( t \right)^2 \left[ d y^2 + d z^2 \right] \;\;\;,\;\;\;
\langle A_\mu \rangle  = \left( 0 ,\, A_1 \left( t \right) ,\, 0 ,\, 0 \right)\nonumber\\
\end{equation}
and we parametrize the two scale factors by
\begin{equation}
a = {\rm e}^{\alpha - 2 \sigma} \;\;,\;\;
b = {\rm e}^{\alpha + \sigma}
\end{equation}
(so that ${\rm e}^\alpha = a^{1/3} \, b^{2/3} \;,\; {\rm e}^\sigma = b^{1/3} \, a^{-1/3}$). Namely $\alpha$ parametrizes the overall volume expansion, while $\sigma$ controls the amount of anisotropy. More accurately, the normalization of the two scale factor is unphysical, and one can shift the values of $\alpha$ and $\sigma$ by a constant factor without changing the physics of the system. Therefore, the 
``initial'' (i.e., at the initial time of a simulation) values $\alpha_{\rm in}$ and $\sigma_{\rm in}$ can be set to any arbitrary value. The degree of anisotropy is related to $\dot{\sigma} / \dot{\alpha}$: a constant $\sigma$ corresponds to a FRW flat geometry, with the Hubble rate given by $\dot{\alpha}$.

The background equation of motion for the vector vev is
\begin{equation}
\ddot{A}_1+\left[ \dot{\alpha} + 4 \dot{\sigma} + 2 \frac{\dot{\phi} \, f' \left( \phi \right)}{f \left(\phi \right)} \right] \dot{A}_1 = 0
\end{equation}
where dot denotes differentiation with respect to time, while $f'$ - and, later,  $V'$ - denotes the derivative of that function with respect to the scalar field. This equation is solved by
\begin{equation}
\dot{A}_1 = p_A \, {\rm e}^{-\alpha-4\sigma}/f \left( \phi \right)^2
\end{equation}
where $p_A$ is constant. We insert this solution in the other nontrivial background equations following from (\ref{action}). These equations then rewrite
\begin{eqnarray}
&&3 \dot{\alpha}^2 - 3 \dot{\sigma}^2 = \frac{1}{M_p^2} \left[
\frac{\dot{\phi}^2}{2} + V \left( \phi \right) + \frac{{\tilde p}_A^2}{2 \, f \left( \phi \right)^2} \right] 
\nonumber\\
&&2 \ddot{\alpha} + 3 \dot{\alpha}^2 + 3 \dot{\sigma}^2 = \frac{1}{M_p^2} \left[ 
- \frac{\dot{\phi}^2}{2} + V \left( \phi \right) - \frac{{\tilde p}_A^2}{6 \, f \left( \phi \right)^2} \right] 
\nonumber\\
&&
\ddot{\sigma} + 3 \dot{\alpha} \, \dot{\sigma} = \frac{{\tilde p}_A^2}{3 \, M_p^2 \, f \left( \phi \right)^2} 
\nonumber\\
&&\ddot{\phi} + 3 \dot{\alpha} \, \dot{\phi} + V' \left( \phi \right) = \frac{{\tilde p}_A^2 \, f' \left( \phi \right)}{f \left( \phi \right)^3}
\label{eq-bck}
\end{eqnarray}
where we have defined
\begin{equation}
{\tilde p}_A \left( t \right) \equiv {\rm e}^{-2\alpha-2\sigma} p_A
\end{equation}

The first of (\ref{eq-bck}) is the $tt$ Einstein equation for the system, and, for 
$\dot{\sigma} = {\tilde p}_A = 0$, it reduces to the standard Friedmann equation. The last term in this equation is the energy density of the vector field. Since the normalization of the scale factors is unphysical, the quantity ${\tilde p}_A$ must be independent of it. Therefore, if one wishes to change the ``initial'' values of the two scale factors, the integration constant $p_A$ must also be changed according to $p_A \propto e^{2 \alpha_{\rm in} + 2 \sigma_{\rm in}}$. The second and fourth equation in (\ref{eq-bck}) are a combination of the spatial $xx$ and $yy=zz$ Einstein equations, while the fourth equation is the equation for the scalar field. One of these three equations can be obtained from the other two, and from the $tt$ Einstein equation, as a consequence of a nontrivial Bianchi identity.

We are interested in an inflationary background solution of (\ref{eq-bck}) in a regime of slow roll and small anisotropy. The first and last of (\ref{eq-bck}) can be approximated as in the standard case, and combined to give $\alpha \left( \phi \right) \approx - \int^\phi  \frac{V}{M_p^2\,V'} \, d \phi$. We can also find how $f$ and $V$ need to be related to each other to have a prolonged anisotropic stage. Namely, we require that the ratio between the energy densities of the vector and the scalar field remains approximately constant during inflation. In the slow roll regime, and for small anisotropy, we have
\begin{equation}
\frac{\rho_A}{\rho_\phi} \approx \frac{p_A^2 \, {\rm e}^{-4\sigma}}{2 \, V \left( \phi \right)} \, \left( \frac{{\rm e}^{-2\alpha}}{f \left( \phi \right)} \right)^2
\label{ratio-rho}
\end{equation}
and, given that the first factor is very slowly evolving in this regime, we require \cite{Watanabe:2009ct} that $f \approx {\rm exp} \left[ - 2 \alpha \right] \approx {\rm exp} \left[  \int^\phi  \frac{2 V}{M_p^2\,V'} \, d \phi \right] \,$. For definiteness, we will consider the simplest chaotic inflationary potential
\begin{equation}
V \left( \phi \right) = \frac{1}{2} \, m^2 \, \phi^2 \;\;\;,\;\;\;
f \left( \phi \right) = {\rm exp} \left( \frac{c \, \phi^2}{2 \, M_p^2} \right)
\label{Vf}
\end{equation}
where $c$ is a numerical constant close to one.

We expect that, for $c>1$, $f$ decreases more quickly than ${\rm e}^{-2\alpha}$ during inflation, so that the anisotropy actually increases; to verify this, we study the first and last of (\ref{eq-bck}) in more details. Namely, we disregard the terms proportional to $\dot{\sigma}$ and ${\tilde p}_A$ in the first equation, and the term proportional to $\ddot{\phi}$ in the last one (it is consistent to keep the term proportional to ${\tilde p}_A$ in the last equation, and disregard it in the first one, provided that $\phi \gg M_p / \sqrt{c}$, which is indeed a good approximation during inflation). These two equations can be then integrated to give \cite{Watanabe:2009ct} 
\begin{equation}
{\rm e}^{c \frac{\phi^2}{M_p^2}+4\alpha} \approx  \frac{c^2}{c-1} \, \frac{p_A^2}{m^2 M_p^2} + D \, {\rm e}^{-4\left( c - 1 \right) \alpha}
\label{slow}
\end{equation}
where $D$ is an integration constant. For $c<1$, the second term on the right hand side of this equation increases over the first one during inflation. If this term is dominant, then Eq. (\ref{slow}) reduces to $\alpha \approx - \phi^2 / \left( 4 \, M_p^2 \right)$ (plus an unphysical constant), which is the standard isotropic result. On the contrary, for $c>1$, the first term of (\ref{slow}) increases over the second term over time. If this term is dominant, Eq. (\ref{slow}) reduces to $\alpha \approx - c \, \phi^2 / \left( 4  \, M_p^2 \right)$. Combining this with the approximate $tt$ Einstein equation, $6 \dot{\alpha}^2 \approx m ^2 \phi^2 /  M_p^2$, gives \cite{Watanabe:2009ct} 
\begin{equation}
3 \, \dot{\alpha} \, \dot{\phi} \approx  - \frac{m^2 \, \phi}{c} \;\;\;\;\;\;\;\;\;\; ( c > 1 )
\label{anisotropic}
\end{equation}
which gives a value for $\dot{\phi}$ about $1/c$ times the standard result. To quantify the anisotropy, we neglect the $\ddot{\sigma}$ term in the third of (\ref{eq-bck}). Combining the resulting expression with (\ref{slow}), in a regime in which the term proportional to $D$ can be disregarded, we find
\begin{equation}
\frac{\dot{\sigma}}{\dot{\alpha}} \approx  \frac{2}{3} \, \frac{c-1}{c^2} \, \frac{M_p^2}{\phi^2} \;\;\;\;\;\;\;\;\;\;  ( c >1 )
\label{attractor}
\end{equation}
which indeed confirms that the anisotropy increases during inflation. 

This quantity is approximately equal the ratio between the energy densities of the vector and the scalar. Indeed, combining eqs. (\ref{ratio-rho}) and
(\ref{slow}), in a regime in which the term proportional to $D$ can be disregarded, we also find
\begin{equation}
\frac{\rho_A}{\rho_\phi} \approx \frac{c-1}{c^2} \, \frac{M_p^2}{\phi^2} \;\;\;\;\;\;\;\;\;\;  ( c >1 )
\label{attractor-rho}
\end{equation}

Namely, for $c<1$ the system evolves towards isotropy, while for $c>1$ the system evolves towards isotropy during inflation. In both cases the solution is an attractive one. One can decide to take $c<1$ and start away from the isotropic attractor solution. If $c$ is sufficiently close to one, the anisotropy will decrease very slowly, and still give some observable nonstandard signature. The result will however be sensitive on the initial conditions, and not only on the model. On the other hand, for $c>1$ we can start in the attractor anisotropic solution characterized by (\ref{anisotropic}). The underlying  idea is that inflation lasted much more than the observable last $60$ e-folds, and that the solution converged to the attractor one during that time. In this case,  the phenomenological signatures of the model are insensitive on the initial conditions, precisely as in the standard inflationary case.~\footnote{Even in the standard case, one can assume that the inflaton was not yet in the attractor solution when the largest observed multipoles left the horizon. For instance, a fast roll evolution at that stage results in a suppression of the CMB quadrupole~\cite{Contaldi:2003zv}. This signal is however dependent on the assumed initial conditions.} For this reason, we only study the $c>1$ case in this work.

We conclude this Section with two remarks. Firstly, we note that the anisotropy is proportional to $c-1$. The anisotropic attractor solution is continuously connected to a FRW solution in the $c\rightarrow 1$ limit. We expect  standard results for the perturbations in this limit, as the computations presented in the next Sections confirm. Secondly, while the anisotropy increases during inflation, it decreases after inflation. Indeed, after inflation $\phi$ oscillates around zero, with a decreasing amplitude. Then $f \rightarrow 1$, and the mechanism of prolonged anisotropy becomes ineffective. The amplitude of the vector rapidly decreases, and the background evolution becomes isotropic.

\begin{figure}
\centering
\includegraphics[width=0.4\textwidth,angle=-90]{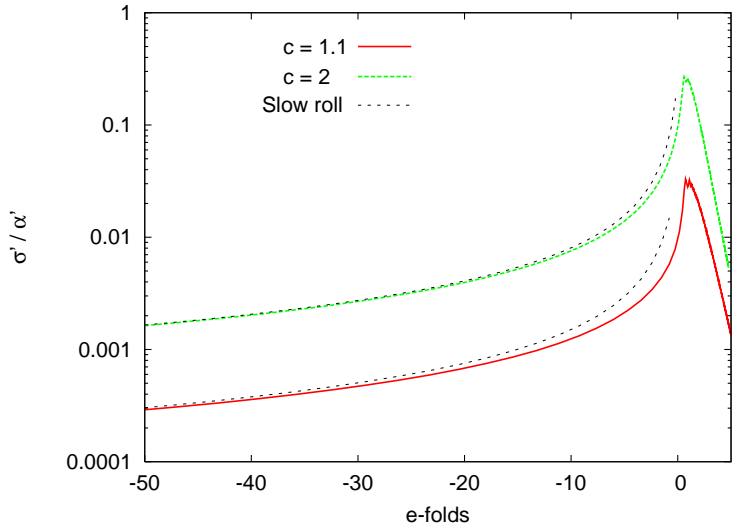}
\caption{
Evolution of the anisotropy factor $\dot{\alpha} / \dot{\sigma}$ for two different values of $c$ in the function (\ref{Vf}), as a function of the number of e-folds. $N=0$ corresponds to the end of inflation.
}\label{fig:bck}
\end{figure}

For illustrative purposes, we show in Figure \ref{fig:bck} the evolution of the anisotropy $\dot{\sigma} / \dot{\alpha}$ as a function of the number of e-folds $N \equiv \alpha$, normalized to zero at the end of inflation.~\footnote{Since $\dot{\sigma} \ll \dot{\alpha}$, we define the number of e-folds $N$ and the end of inflation as in the FRW case. This gives $N=\alpha$, and the end of inflation occurs when $\ddot{\alpha} + \dot{\alpha}^2$ becomes negative.} We show the evolution for two different values of $c$, starting from the slow roll anisotropic initial condition.~\footnote{Specifically, we use the last three equations of (\ref{eq-bck}) in our numerical evolutions. We also satisfy the first of (\ref{eq-bck}) through the initial conditions (if this equation holds at the initial time, it is preserved by the remaining equations): we relate the initial value of $\dot{\phi}$ and of $\dot{\sigma}$ to that of $\dot{\alpha}$ through eqs. (\ref{anisotropic}) and (\ref{attractor}); we insert these expressions in the first of  (\ref{eq-bck}), and we then use this equation to relate $\dot{\alpha}_{\rm in}$ to the initial value of $\phi$. In this way, all the initial conditions are given in terms of $\phi_{\rm in}$, which completely specifies any point along the attractor solution.} We also show the evolution of the anisotropy parameter given by the slow roll solution, eq. (\ref{attractor}). We see that the slow roll expression is very accurate during most of the inflationary evolution.

\section{Perturbations}

This Section studies the perturbations of the model \cite{Watanabe:2009ct} discussed in the previous Section. The discussion is divided in several Subsections. In Subsection \ref{subsec:class} we introduce the perturbations, and we classify them according to how they behave with respect to $2$d spatial rotations in the isotropic $yz$ plane. We also perform the gauge choices which completely fix the freedom associated with general coordinate transformations, and with the U(1) invariance of the vector. We then integrate out the nondynamical perturbations, and we provide the quadratic action for the dynamical modes. In Subsection \ref{subsec:quant} we quantize this action, and we define the initial adiabatic vacuum, valid during inflation in the sub-horizon regime. In Subsection \ref{subsec:twopoint} we introduce the two point correlation functions for anisotropic backgrounds, and we express them in terms of angular dependent power spectra. In Subsection \ref{subsec-eom-in} we write more explicitly the initial conditions for the modes (coming from the adiabatic vacuum obtained in \ref{subsec:quant}) and their evolution equations (coming from the action obtained in \ref{subsec:class}). Finally, in Subsection  
\ref{subsec:spectra} we discuss which combination of perturbations reduce to the standard one as the universe isotropizes, and we provide their power spectra.

Both the computation of the quadratic action, and its quantization, follow formalisms that have been developed elsewhere. Although we made an effort to keep the present discussion self contained, some details have been omitted here for brevity. The interested reader can refer to appendix B of \cite{Gumrukcuoglu:2007bx} and to Section III of \cite{Himmetoglu:2009qi} for more details on the computation of the quadratic action, and to Section II of \cite{Nilles:2001fg} for the quantization of the perturbations based on this action.

\subsection{Classification, gauge choice, and quadratic action for the dynamical modes} 
\label{subsec:class}

The most general set of perturbations about the anisotropic background solution discussed in the previous Section is given by
\begin{equation}
\delta g_{\mu \nu} = \left( \begin{array}{cccc}
- 2 \Phi & a \, \partial_x \chi & b \left( \partial_i B + B_i \right) \\
& - 2 a^2 \Psi & a b \partial_x \left( \partial_i {\tilde B} + {\tilde B}_i \right) \\
& & b^2 \left[ -2 \Sigma \delta_{ij} + 2 E_{,ij} + E_{(i,j)} \right] 
\end{array} \right) \;\;\;,\;\;\;
\delta A_\mu = \left( \delta A_0 ,\, \delta A_1 ,\, \partial_i \delta A +  \delta A_i \right)
\;\;\;,\;\;\; \delta \phi
\label{perts-def1}
\end{equation}
where the index $i=1,2$ spans the isotropic $yz$ plane, and $E_{(i,j)} \equiv \partial_i E_j + \partial_j E_i$. With this choice, the perturbations are decomposed in scalar and vector with respect to rotations in the isotropic plane. This is a convenient procedure since the two different types of modes are decoupled from each other at the linearized level, and can be studied separately 
\cite{Gumrukcuoglu:2007bx}.~\footnote{This strategy is mutuated from the one adopted in the standard $3$ dimensional case, in which the modes are classified in scalar, vector and tensors with respect to $3$d rotations. Two dimensional tensor modes do not exist, since imposing the transversality and traceless conditions eliminates all the degrees of freedom of a $2\times2$ symmetric tensor.} Specifically, the modes $B_i ,\, {\tilde B}_i \, E_i ,\, \delta A_i$ are $2$d vector modes, satisfying $\partial_i B_i = \dots = 0$, and they encode one degree of freedom each. The remaining modes are $2$d scalars, which also encode one degree of freedoms each. Altogether, we have $15$ degrees of freedom in the perturbations (\ref{perts-def1}).

We Fourier transform the perturbations,
\begin{equation}
\delta \left( t ,\, {\bf x} \right) = \int \frac{d^3 k}{\left( 2 \pi \right)^{3/2}} \, {\rm e}^{i \, {\bf k} \cdot {\bf x}} \, \delta \left( t ,\, {\bf k} \right)
\label{ft}
\end{equation}
where $\delta$ denotes any of the perturbations, and we use the same symbol for a perturbation in 
real and in momentum space. The reality of $\delta \left( {\bf x} \right)$ imposes the condition $\delta^\dagger \left( {\bf k} \right) = \delta \left( {\bf - k} \right)$ (we use hermitian conjugate, and not simply charge conjugate, since in the next Section the modes are treated as operators, in order to quantize them). At the linearized level, all modes are decoupled from each other, and can be studied separately. We denote the comoving momentum of the mode as ${\bf k} = \left( k_L ,\, k_{T2} ,\, k_{T3} \right)$. Due to the symmetry in the $y-z$ plane, physical results depend on $k_T \equiv \sqrt{k_{T2}^2 + k_{T3}^2}$ rather than on $k_{T2}$ and $k_{T3}$ separately.~\footnote{This was exploited in \cite{Gumrukcuoglu:2007bx}, where $k_{T3}$ was set to zero. As a consequence, the perturbations in real space do not depend on $z$, and the decomposition (\ref{perts-def1}) was simpler. Here we adopt the more general formulation used in \cite{Gumrukcuoglu:2008gi}, where both $k_{T,2}$ and $k_{T,3}$ can be nonvanishing. We also point out that spatial derivatives are introduced in the parametrization (\ref{perts-def1}) for algebraic convenience. In practice, we assume that both $k_L$ and $k_T$ are nonvanishing. This restriction does not affect the computation of any observable, since modes with $k_L=0$ and/or $k_T=0$ constitute a subset of zero measure when one integrates over the momentum to obtain results in real space.} We denote the physical momentum of the mode by ${\bf p} = \left( p_L ,\, p_{T2} ,\, p_{T3} \right)$, where $p_L = k_L / a$ and $p_{Ti} = k_{Ti}/b$. 
As for the comoving momentum, we  define $p_T = \sqrt{p_{T2}^2 + p_{T3}^2}$. 

It is convenient to write write explicitly the single degree of freedom encoded in the $2$d vector modes. In momentum space, we have
\begin{equation}
B_i \equiv i \, \epsilon_{ij} k_{Tj} \, B_v \;\;\;,\;\;\;
{\tilde B}_i \equiv i \, \epsilon_{ij} k_{Tj} \, {\tilde B}_v \;\;\;,\;\;\;
E_i \equiv i \, \epsilon_{ij} k_{Tj} \, E_v \;\;\;,\;\;\;
\delta A_i \equiv i \, \epsilon_{ij} k_{Tj} \, \delta A_v
\label{2dv-transv}
\end{equation}
where $\epsilon_{ij}$ is antisymmetric, and $\epsilon_{12}=1$.

To proceed, we need to fix the gauge freedoms of the system. We start from the freedom associated with general coordinate transformations. To do so, we can either choose a gauge that completely removes this freedom, as done in \cite{Gumrukcuoglu:2007bx,Gumrukcuoglu:2008gi}, or we can rewrite the action and the equations for the perturbations in terms of gauge invariant modes, as done in 
\cite{Himmetoglu:2008hx,Himmetoglu:2009qi}. The two procedures are equivalent. In the present work, we choose the first one, which is algebraically simpler. Specifically, we set
\begin{equation}
\delta g_{1i,{\rm 2ds}} = \delta g_{ij} = 0
\label{ourgauge}
\end{equation}
which, in the parametrization (\ref{perts-def1}), gives ${\tilde B} = \Sigma = E = E_i = 0$. One can check (see Appendix B.1 of \cite{Gumrukcuoglu:2007bx}) that indeed (i) this choices can be always made, and (ii) it fixes completely the freedom associated with thee coordinate transformations. There is also the U(1) gauge associated with the transformations $A_\mu \rightarrow A_\mu + \partial_\mu \xi$. We fix this by setting $\delta A = 0$. 

This gauge fixing leaves us with seven $2$ds modes ($\Phi ,\, \chi ,\, B ,\, \Psi ,\, \delta A_0 ,\, \delta A_1 ,\, \delta \phi $) and three $2$dv modes ($B_i ,\, {\tilde B}_i ,\, \delta A_i $). Not all these modes correspond to physically propagating degrees of freedom. Indeed the modes $\delta A_0$ and $\delta g_{0\mu}$ enter in the quadratic action of the perturbations without time derivatives~\cite{Gumrukcuoglu:2007bx}. As a consequence, the equations of motion for the perturbations are algebraic in them, and their value is specified in terms of the values of the other modes, without introducing independent degrees of freedom. Our gauge choice is motivated by the fact that it preserves the $\delta A_0$ and $\delta g_{0\mu}$ perturbations, so that the identification of the nondynamical modes is immediate (in other gauges, the nondynamical modes correspond to more complicated linear combinations of the perturbations which are preserved in those gauges). The nondynamical modes need to be integrated out of the action.~\footnote{This is what is also done in the standard computations
\cite{Mukhanov:1990me}. For instance, in the case of a single scalar inflaton on a FRW background, the number of perturbations is $11$ (initial perturbations in the metric and in the scalar) $-4$ (after gauge fixing) $-4$ (after eliminating the nondynamical modes) $=3$ (namely, the scalar density contrast, and the two gravity waves polarization). Our procedure extends this computation to the more general background we are studying. In Appendix \ref{appendix-iso} we verify that our computation reduces to the standard one in the limit of isotropic background.} Namely, we express them in terms of the dynamical modes (through the corresponding Einstein equations), and we insert these expressions back into the action. In this way, we are left with an action in terms of the dynamical modes only~\footnote{The procedure of integrating out the nondynamical modes is described in details in Section III of \cite{Himmetoglu:2009qi}.} 
\begin{equation}
S^{(2)}_{\rm perts.} = S_{\rm 2ds}^{(2)}  \left[ \psi,\, \delta \phi ,\, \delta A_1 \right] +  S_{\rm 2dv}^{(2)} \left[ {\tilde B}_i ,\, \delta A_i \right]
\label{action-perts}
\end{equation}
The fields entering in these actions are not canonically normalized. The canonically normalized fields are obtained through the redefinitions
\begin{eqnarray}
\delta \phi &\equiv& {\rm e}^{-\frac{3}{2} \alpha} \left(  V_+ - \frac{\dot{\phi}}{\sqrt{2} M_p \left( \dot{\alpha} + \dot{\sigma} \right)} \, H_+ \right)
\nonumber\\
\Psi &\equiv& {\rm e}^{-\frac{3}{2} \alpha} \, \frac{2 p^2 \dot{\alpha} + \left( 2 p_L^2 - p_T^2 \right) \dot{\sigma}}{\sqrt{2} \, M_p \, p_T^2 \, \left( \dot{\alpha} + \dot{\sigma} \right)} \, H_+
\nonumber\\
\alpha_1 &\equiv& {\rm e}^{- \frac{1}{2} \alpha - 2 \sigma} \left[
\frac{ p}{f \left( \phi \right) p_T} \, \Delta_+ - \frac{{\tilde p}_A}{\sqrt{2} M_p \, f \left( \phi \right)^2 \left( \dot{\alpha} + \dot{\sigma} \right)} H_+ \right]
\label{2ds-can}
\end{eqnarray}
in the $2$ds sector, and
\begin{eqnarray}
{\tilde B}_i &\equiv&  - \sqrt{2} \epsilon_{ij} \, k_{Tj} \, {\rm e}^{-\frac{3 \alpha}{2}} \, \frac{\sqrt{{\rm e}^{6 \sigma} k_L^2 + k_T^2}}{M_p \, k_L \, k_T^2} \, H_\times \nonumber\\
\delta A_i &\equiv&  i \, \epsilon_{ij} \, \frac{k_{Tj}}{k_T} \, \frac{ {\rm e}^{\sigma-\frac{\alpha}{2}}}{f \left( \phi \right)} \, \Delta_\times
\label{2dv-can}
\end{eqnarray}
in the $2$dv sector (where $\epsilon_{ij}$ is antisymmetric, and $\epsilon_{12}=1$).

Once expressed in terms of these fields, the two actions in (\ref{action-perts}) rewrite 
\begin{eqnarray}
S^{(2)}_{2ds} &=& \frac{1}{2} \int d t \, d^3 k \left[ \dot{Y_s}^\dagger \dot{Y_s} + \dot{Y}^\dagger K_s Y_s
- Y_s^\dagger K_s \dot{Y_s} - Y^\dagger \, \Omega_s^2 \, Y_s \right] \nonumber\\
S^{(2)}_{2dv} &=& \frac{1}{2} \int d t \, d^3 k \left[ \dot{Y_v}^\dagger \dot{Y_v} + \dot{Y_v}^\dagger K_v \, Y_v - Y_v^\dagger K_v \dot{Y_v} - Y_v^\dagger \, \Omega_v^2 \, Y_v \right] 
\label{action-formal}
\end{eqnarray}
where
\begin{equation}
Y_s \equiv \left( \begin{array}{c} V_+ \\ H_+ \\ \Delta_+ \end{array} \right) \;\;\;,\;\;\;
Y_v \equiv \left( \begin{array}{c} H_\times \\ \Delta_\times \end{array} \right) 
\end{equation}
and where the explicit expressions for the matrices $K_{s,v}$ and $\Omega^2_{s,v}$ are given in 
Appendix \ref{appendix-explicit}.

\subsection{Quantization and initial adiabatic vacuum} \label{subsec:quant}

We need to quantize the two actions (\ref{action-formal}) in order to provide the initial conditions for the modes and the expressions for the correlators. We do not need to discuss the two systems separately, since the two actions are formally the same, and the fields and the matrices entering in them have identical properties. In both cases, the actions are of the type
\begin{equation}
S = \frac{1}{2} \int d t \, d^3 k \left[ \dot{Y}^\dagger \dot{Y} + \dot{Y}^\dagger \, K \, Y
- Y^\dagger \, K \, \dot{Y} - Y^\dagger \, \Omega^2 \, Y \right] 
\label{act-Y}
\end{equation}
where $Y$ is an array of fields, $K$ a real and anti-symmetric matrix, and $\Omega^2$ a real and symmetric matrix. These matrices are unchanged under the parity transformation ${\bf k} \rightarrow {\bf - k}$ (as can be  seen from the explicit expressions given in Appendix \ref{appendix-explicit}). It can be checked from the reality condition stated after eq. (\ref{ft}), and from the definitions (\ref{2ds-can}) and (\ref{2dv-can}) of the canonical modes, that, for both systems, any of the $Y_i$ fields entering in the array $Y$ satisfies $Y_i^\dagger \left( {\bf k} \right) = Y_i \left( - {\bf k} \right) \,$.

To remove the mixed terms proportional to the matrix $K$, we first perform the field redefinition 
\begin{equation}
\psi \equiv R \, Y
\label{Ypsi}
\end{equation}
where $R$ is an orthogonal matrix (so that $\dot{Y}^\dagger \dot{Y} = \dot{\psi}^\dagger \dot{\psi} \,$), satisfying 
\begin{equation}
\dot{R} = R \, K \;\;\;,\;\;\; R_{\rm late} = \identity
\end{equation}
where the second condition states that $R$ should reduce to the identity at late times, when the universe becomes isotropic, and $K\rightarrow 0$ (in which case, the rotation (\ref{Ypsi}) is no longer needed).
The first condition can be also written as $K = R^T \, \dot{R}$; we also note that $R \left( {\bf - k} \right) = R \left( {\bf k} \right)$, since this property is also satisfied by $K$. As a consequence, each of the fields entering in the array $\psi$ satisfies $\psi_i^\dagger \left( {\bf k} \right) = \psi_i \left( - {\bf k} \right) \,$. 
In terms of the fields $\psi$, the action (\ref{act-Y}) rewrites
\begin{equation}
S = \frac{1}{2} \int d t \, d^3 k \left[ \dot{\psi}^\dagger \dot{\psi} - \psi^\dagger \, {\tilde \Omega}^2 \, \psi \right]  \;\;\;,\;\;\; {\tilde \Omega}^2 \equiv R \left( \Omega^2 + K^T \, K \right) R^T
\label{act-psi}
\end{equation}
where we note that ${\tilde \Omega}^2$ is real, symmetric, and invariant under ${\bf k} \rightarrow {\bf - k}$. 

We can also define the real space fields $\psi \left( t ,\, x \right)$ as in (\ref{ft}). These fields are real, and their action is formally identical to the action for the coupled bosonic system quantized in 
\cite{Nilles:2001fg}. Therefore, we quantize the fields $\psi_i$ as done in that work.

We introduce the matrix $C$ satisfying
\begin{eqnarray}
&&C^T \, {\tilde \Omega}^2 \,  C = {\rm diag } \left( \omega_1^2 ,\, \dots ,\, \omega_N^2 \right) \equiv \omega^2 \nonumber\\
&&C_{\rm end} = \identity
\label{matC}
\end{eqnarray}
where the second condition follows from the fact that, in the late time isotropic limit ${\tilde \Omega}^2 = \Omega^2$ is already diagonal. We note that  $C$ is orthogonal, and unchanged under ${\bf k} \rightarrow {\bf - k}$. We then define
\cite{Nilles:2001fg}
\begin{eqnarray}
\psi_i \left( {\bf k} \right) &=& C_{ij} \left[ h_{jl} \left( {\bf k} \right)\, {\hat a}_l \left( {\bf k} \right)+ 
h_{jl}^* \left( {\bf k} \right) \, {\hat a}_l^\dagger \left( {\bf - k } \right) \right] \nonumber\\
\pi_i \left( {\bf k} \right) = \dot{\psi}_i \left( {\bf k} \right)
&=& C_{ij} \left[ {\tilde h}_{jl} \left( {\bf k} \right)\, {\hat a}_l \left( {\bf k} \right)+ 
{\tilde h}_{jl}^* \left( {\bf k} \right) \, {\hat a}_l^\dagger \left( {\bf - k } \right) \right] 
\label{quant}
\end{eqnarray}
where ${\hat a}$ and ${\hat a}^\dagger$ are annihilation and creation operators, respectively,
satisfying
\begin{equation}
\left[ a_i \left( {\bf k} \right) ,\, a_j^\dagger \left( {\bf k'} \right) \right] = \delta^3 \left( {\bf k} - {\bf k'} \right) \, \delta_{ij}
\label{a-ad}
\end{equation}
From the equations of motion following from (\ref{act-psi}), and from the fact that $\pi_i = \dot{\psi}_i$, we find that the coefficients $h_{ij}$ and ${\tilde h}_{ij}$, obey the evolution equations (in matrix from)
\begin{equation}
\dot{h} = {\tilde h} - \Gamma \, h \;\;\;,\;\;\;
\dot{\tilde h} = - \Gamma \, {\tilde h} - \omega^2 h
\;\;\;,\;\;\; \Gamma \equiv C^T \, \dot{C}
\label{eom-hht}
\end{equation}
From the parity properties of the matrices $C$ and $\omega$, and from the initial conditions (which we determine below, see eq. (\ref{early})), we see that $h_{ij}$ and ${\tilde h}_{ij}$ are unchanged under ${\bf k} \rightarrow {\bf -k}$.~\footnote{This is why we wrote $h_{ij}^* \left( {\bf k} \right)$ and ${\tilde h}_{ij}^* \left( {\bf k} \right)$, rather than 
$h_{ij}^* \left( {\bf -k} \right)$ and ${\tilde h}_{ij}^* \left( {\bf -k} \right)$, in the decompositions (\ref{quant}).}

We further define 
\begin{eqnarray}
h = \frac{1}{\sqrt{2 \omega}} \, \left( \alpha + \beta \right) \;\;,\;\;
{\tilde h} = \frac{-i \omega}{\sqrt{2 \omega}} \, \left( \alpha - \beta \right) 
\label{def-ab}
\end{eqnarray}
It has been shown in \cite{Nilles:2001fg} that the normal ordered hamiltonian for the fields $\psi_i$ can be then cast in the form
\begin{equation}
{\hat H} = \int d^3 k \, \omega_i \, {\hat b}_i^\dagger \left( {\bf  k} \right) \, \, {\hat b} \left( {\bf k} \right)
\label{ham-diag}
\end{equation}
where ${\hat b}_i$ and ${\hat b}_i^\dagger$ are new annihilation and creation operators, related to those defined in (\ref{quant}) by (notice that the matrices $\alpha$ and $\beta$ are unchanged under
${\bf k} \rightarrow {\bf - k}$)
\begin{equation}
\left( \begin{array}{c} 
{\hat b} \left( t ,\, {\bf k} \right) \\ {\hat b}^\dagger \left( t ,\, {\bf - k} \right)
\end{array} \right) \equiv
\left( \begin{array}{cc}
\alpha & \beta^* \\
\beta & \alpha^* 
\end{array} \right)_{t,\,{\bf k}}
\, 
\left( \begin{array}{c} 
{\hat a} \left( {\bf k} \right) \\ {\hat a}^\dagger \left( {\bf - k} \right)
\end{array} \right) 
\end{equation}

We see that the hamiltonian is diagonal in the ${\hat b}_i , {\hat b}_i^\dagger$ basis, so that these operators annihilate and create quanta of the (time-dependent) physical eigenstates of the system.
The matrices $\alpha$ and $\beta$ generalize to a system of $N$ coupled fields the Bogolyubov coefficients that are needed for the quantization of a field with time dependent frequency. As shown in
\cite{Nilles:2001fg}, the canonical quantization of the $\psi_i$ fields imposes the conditions
\begin{equation}
\alpha \alpha^\dagger - \beta^* \beta^T = {\identity} \;\;\;\;,\;\;\;\;
\alpha \beta^\dagger - \beta^* \alpha^T =0
\label{cond-ab}
\end{equation}
Moreover, from the evolution equations (\ref{eom-hht}), and the definitions (\ref{def-ab}), one finds that $\alpha$ and $\beta$ obey the evolution equations \cite{Nilles:2001fg}
\begin{equation}
\left\{ \begin{array}{l}
\dot{\alpha} = - i \omega \alpha + \frac{\dot{\omega}}{2 \omega} \beta - I \alpha - J \beta \\ \\
\dot{\beta} = i \omega \beta + \frac{\dot{\omega}}{2 \omega} \alpha - I \beta - J \alpha 
\end{array} \right.
\label{eom-ab}
\end{equation}
where 
\begin{equation}
I = \frac{1}{2} \left( \sqrt{\omega} \Gamma \frac{1}{\sqrt{\omega}} + \frac{1}{\sqrt{\omega}} \Gamma \sqrt{\omega} \right) \;\;\;,\;\;\; 
J = \frac{1}{2} \left( \sqrt{\omega} \Gamma \frac{1}{\sqrt{\omega}} - \frac{1}{\sqrt{\omega}} \Gamma \sqrt{\omega} \right) 
\end{equation}

An inspection of the initial matrices $K$ and $\Omega^2$ shows that, at early times (when the mode is deeply inside the horizon) $\Omega^2 = p^2 \, \identity + {\rm O} \left( H \right)$ and 
$K = {\rm O } \left( \sqrt{c-1} \, H \right) \,$ (in these expressions, $H \equiv \dot{a}/a \simeq \dot{b} / b$ under the assumption of small anisotropy that we are making in this work). As a consequence, 
\begin{equation}
\omega_i^2 \simeq p^2 + {\rm O } \left( H^2 \right) \;\;\;\;\;,\;\;\;\;\;
\Gamma,\, I ,\, J ,\, \frac{\dot{\omega}}{\omega} = {\rm O } \left( H \right)
\end{equation}
in this early time regime. Therefore, we can disregard all but the first term in both of the right hand sides of (\ref{eom-ab}). This leads us to the initial adiabatic vacuum solutions~\footnote{Notice that 
$\alpha_{\rm early}$ is diagonal; the allowed initial conditions are actually more general than (\ref{early}), since one can multiply each diagonal entry of $\alpha_{\rm early}$ by a constant, and arbitrary, 
 phase factor ${\rm e}^{i \gamma_i} $. This amounts in changing the matrices $\alpha$ and $\beta$ given here by a matrix multiplication from the right, $\alpha \rightarrow \alpha \, P ,\, \beta \rightarrow \beta \, P$, where $P \equiv {\rm diag } \left( {\rm e}^{i \gamma_1} ,\, \dots {\rm e}^{i \gamma_N} \right)$. The equations of motion (\ref{eom-ab}) are unchanged by this multiplication. The same is true in terms of the matrices $h$ and ${\tilde h}$. Since $h$ enters in the observable two point correlation function through the combination $h \, h^\dagger$, see eq. (\ref{2point}), the matrix $P$ drops from the observable result. This confirms that the arbitrary phases contained in $P$ are unphysical, and can be set to any value. We use this freedom to set all the phases to zero at the initial time of our numerical simulations.}
\begin{equation}
\alpha_{\rm early} = {\rm e}^{-i \int^t d t \, \omega }  
\;\;\;,\;\;\;
\beta_{\rm early} = 0
\label{early}
\end{equation}

This solution obeys (\ref{eom-ab}) in the early time regime, satisfies the quantization conditions (\ref{cond-ab}), and corresponds to an initial empty vacuum, if we impose that the vacuum state is annihilated by the operators ${\hat a}_i$ entering in  (\ref{quant}).

\subsection{Two point correlation functions} \label{subsec:twopoint}

By combining the various redefinitions given in the above Section, we can write the canonically normalized fields in real space as
\begin{equation}
Y_i \left( t ,\, {\bf x} \right) = \int \frac{d^3 k}{\left( 2 \pi \right)^{3/2}} \, {\rm e}^{i {\bf k} \cdot {\bf x}} \, \left[ 
\Upsilon_{ij} \left( t,\, {\bf k} \right) \, {\hat a}_j \left( {\bf k} \right) + \Upsilon_{ij}^* \left( t,\, {\bf k} \right) \, {\hat a}_j^\dagger \left( {\bf - k} \right) \right]
\label{combine}
\end{equation}
where
\begin{equation}
\Upsilon_{ij} \left( {\bf k} \right) \equiv \left( R^T \, C \, h \right)_{ij} 
\label{upsilon}
\end{equation}
(notice that $h$ and $\Upsilon$ coincide at late times).

The (statistically averaged) two point correlation function can be expressed as the quantum expectation value
\begin{equation}
{\cal C}_{ij} \left( {\bf x} ,\, {\bf y} \right) \equiv \frac{1}{2} \langle Y_i \left( t ,\, {\bf x} \right) \, 
Y_j \left( t ,\, {\bf y} \right) + Y_j \left( t ,\, {\bf y} \right) \, Y_i \left( t ,\, {\bf x} \right) \rangle
\label{2point-def}
\end{equation}
where the symmetrization is required since the statistical average is a classical operation, independent of the ordering chosen (when computing any correlation with real data, ${\cal C}_{ij} \left( {\bf x} ,\, {\bf y} \right) = {\cal C}_{ji} \left( {\bf y} ,\, {\bf x} \right)$).

We insert (\ref{combine}) into (\ref{2point-def}). The resulting expression can be then simplified using 
the commutation relations (\ref{a-ad}) and the fact that the vacuum state is annihilated by ${\hat a}_i$ at all times (since the quantization of the previous Subsection is performed in the Heisenberg picture). After some algebra, we find
\begin{equation}
{\cal C}_{ij} \left( {\bf x} ,\, {\bf y} \right) = \int \frac{d^3 k}{\left( 2 \pi \right)^3} \, {\rm e}^{i {\bf k} \cdot \left( {\bf x} - {\bf y} \right)} \, {\rm Re \, } \left[ \left( \Upsilon \, \Upsilon^\dagger  \right)_{ij}  \right] 
\label{2point}
\end{equation}

We now define the power spectra associated with these correlators. All of them are of the type
\begin{equation}
{\cal C}_{\cal F} = \int \frac{d^3 k}{\left( 2 \pi \right)^3} \, {\rm e}^{i {\bf k} \cdot \left( {\bf x} - {\bf y} \right)}  {\cal F} \left( {\bf k} \right)
\label{corr-f}
\end{equation}
where the function ${\cal F}$ is real and it depends only on the absolute values of the components of ${\bf k}$ along the anisotropic $x-$direction (which we denoted by $\vert k_L \vert$), and on the $y-z$ plane (denoted by $k_T = \sqrt{k_{T2}^2+k_{T3}^2}$). Thanks to this property (which follows from the symmetry of the background under rotations in the $yz$ plane, and under parity), for any two points ${\bf x}$ and ${\bf y}$, we can always choose the $y$ and $z$ axes of the system such that the third component of ${\bf x} - {\bf y}$ vanishes (without changing ${\cal F}$). We therefore set
\begin{equation}
{\bf r} \equiv {\bf x} - {\bf y} \equiv \left( r_L ,\, r_T ,\, 0 \right)
\end{equation}
and ${\bf k} = k \left( \xi ,\, \sqrt{1-\xi^2} \, \cos{\phi_k} ,\, \sqrt{1-\xi^2} \, \sin{\phi_k} \right)$ in the integral
(\ref{corr-f}), where $\xi$ is the cosine of the angle between the $x-$axis and ${\bf k}$. The function ${\cal F}$ does not depend on $\phi_k$ and is even in $\xi$. The integral over $\phi_k$ then gives
\begin{equation}
{\cal C}_{\cal F} = \int \frac{d k}{k} \,  \int_0^1 d \xi \, \cos \left( k \, \xi \, r_L \right)
J_0 \left( k \, \sqrt{1-\xi^2} \, r_T \right) \, P_{\cal F} 
\label{corr-aniso}
\end{equation}
where $J$ is the Bessel function of the first kind, and where we have introduced the power spectrum
\begin{equation}
P_{\cal F} \equiv \frac{k^3}{2 \pi^2} \, {\cal F} \left( k ,\, \xi \right)
\label{powerspectrum}
\end{equation}
In the case at hand, the power spectra depend both on the magnitude of the momentum of the modes, and on the angle between the momentum and the anisotropic direction. 

On anisotropic backgrounds, the power spectrum is isotropic, and eq. (\ref{corr-aniso}) reduces to the standard expression
\begin{equation}
{\cal C}_{\cal F} = \int \frac{d k}{k} \, \frac{\sin \left( k \, r \right)}{k \, r} \, {\cal P}_{\cal F}
\;\;\;\;\;\;\;\;\;\;\;\;
{\rm isotropy}
\end{equation}

\subsection{Evolution of the perturbations and initial conditions} \label{subsec-eom-in}

The equations of motion for the dynamical perturbations follow from (\ref{action-formal}). We expressed the action in momentum space, by Fourier transforming the starting modes as in (\ref{ft}), and by introducing the canonically normalized fields in (\ref{2ds-can}) and (\ref{2dv-can}). The equations of motion for the coefficient of the Fourier transforms of the canonical fields are
\begin{equation}
\ddot{Y_s} + 2 \, K_s \, \dot{Y_s} + \left( \Omega_s^2 + \dot{K_s} \right) Y_s = 0 \;\;\;,\;\;\;
\ddot{Y_v} + 2 \, K_v \, \dot{Y_v} + \left( \Omega_v^2 + \dot{K_v} \right) Y_v = 0
\label{eom-formal}
\end{equation}
where the matrices $K_{s,v}$ and $\Omega_{s,v}^2$ are given in Appendix \ref{appendix-explicit}.

We stress that these equations are a closed subset of the linearized Einstein equations for the perturbations. Specifically, all the perturbations of the model can be divided in dynamical and nondynamical ones. The nondynamical ones enter in the Einstein equations without time derivatives. One can solve the Einstein equations for these perturbations, and express the latter in terms of the dynamical ones. One then insert these expressions into the remaining Einstein equations. The resulting expressions coincide with eqs. (\ref{eom-formal}).~\footnote{The explicit proof of this is given in Section III of \cite{Himmetoglu:2009qi} for details.}

We need to exactly specify what the coefficients $Y_{s,i}$ and $Y_{v,i}$ exactly are. The standard way to compute the generation of perturbations during inflation is a semiclassical computation, in which the perturbations are quantum fields on a classical background \cite{Mukhanov:1990me}. We take this approach in this work. In most systems, the canonical perturbations are decoupled from each other, and can be quantized separately. This is not the case for the system we are studying, and the quantization had to be done accordingly. In Subsection \ref{subsec:quant}, we introduced an array of annihilation/creation operators, and we saw that the Fourier coefficients of (\ref{ft}) are actually linear combinations of these operators, cf. eq. (\ref{combine}):
\begin{equation}
Y_i \left( t ,\, {\bf k} \right) = \Upsilon_{ij} \left( t,\, {\bf k} \right) \, {\hat a}_j \left( {\bf k} \right) + \Upsilon_{ij}^* \left( t,\, {\bf k} \right) \, {\hat a}_j^\dagger \left( {\bf - k} \right)
\end{equation}
Inserting this decomposition into (\ref{eom-formal}), we find
\begin{equation}
\left[ \ddot{\Upsilon}_{ij} + 2 K_{il} \dot{\Upsilon}_{lj} + \left( \Omega^2 + \dot{K} \right)_{il} \, \Upsilon_{lj} \right] a_j + \left[ \ddot{\Upsilon}_{ij}^* + 2 K_{il} \dot{\Upsilon}_{lj}^* + \left( \Omega^2 + \dot{K} \right)_{il} \, \Upsilon_{lj}^* \right] a_j^\dagger = 0 
\end{equation}
both in the $2$d scalar and $2$d vector sector. The linear combinations multiplying different annihilation and creation operators need to cancel separately; therefore
\begin{equation}
\ddot{\Upsilon}_{ij} + 2 K_{il} \dot{\Upsilon}_{lj} + \left( \Omega^2 + \dot{K} \right)_{il} \, \Upsilon_{lj} = 0
\label{eom-upsilon}
\end{equation}
These equations also guarantee that the linear combinations multiplying the annihilation operators vanish, since the matrices $K$ and $\Omega^2$ are real.

Having determined the evolution equations obeyed by $\Upsilon_{ij}$, we now turn to the determination of their initial condition. We find
\begin{equation}
\Upsilon = R^T \, C \, h \;\;\;,\;\;\;
\dot{\Upsilon} = - K \, R^T \, C \, h + R^T \, C \, {\tilde h}
\end{equation}
The first expression is simply the definition (\ref{upsilon}). The second expression is obtained by differentiating the first one, and by using the first and third of (\ref{eom-hht}), as well as $K = - \dot{R}^T R \,$.

The initial values of $R^T \, C$ and $\omega$ are obtained by the diagonalization of $\Omega^2 + K^T K$ at the initial time. Indeed $R^T \, C$ is defined as the matrix that diagonalizes $\Omega^2 + K^T K$, while $\omega^2$ is the diagonal matrix formed by the eigenvalues of $\Omega^2 + K^T K$, see eqs. (\ref{act-psi}) and (\ref{matC}). The initial values of $h$ and ${\tilde h}$ follow instead from eqs. (\ref{def-ab}), and from the initial conditions for $\alpha$ and $\beta$ according to the adiabatic vacuum prescription, eq. (\ref{early}): $\alpha_{\rm in} = \identity ,\, \beta_{\rm in} = 0$.

\subsection{Power spectra after isotropization}
\label{subsec:spectra}

As shown in Appendix~\ref{appendix-iso}, as the universe becomes isotropic after inflation, the canonical perturbations that we have introduced in \ref{subsec:class} become the standard scalar and tensor modes of FRW cosmology. More precisely, we find that, in this regime,
\begin{equation}
R = \frac{H}{a^{3/2} \, \dot{\phi}} \, V_+ \;\;\;,\;\;\;
h_+ = - \frac{\sqrt{2}}{a^{3/2} \, M_p} \, H_+ \;\;\;,\;\;\;
h_\times = \frac{i \, \sqrt{2}}{a^{3/2} \, M_p} \, H_\times
\label{iso-gaugeinv}
\end{equation}
where ${\cal R}$ is the standard (scalar) comoving curvature perturbation, and  $h_+$ and $h_\times$ are the two standard gravity wave polarizations. 

We also shown in Appendix~\ref{appendix-iso} that, in the isotropic regime, our formalism reproduces the standard evolution equations for these modes. These equations are decoupled, so that the modes evolve independently from each other (at the linearized level). However, the two $2$d scalar modes  $V_+$ and $H_+$ are coupled to each other, and to the mode $\Delta_+$ during inflation, when the background is anisotropic. The mode $H_\times$ is always decoupled from these three modes, but it is coupled to the mode $\Delta_\times$ during inflation. The coupling modifies the diagonal correlation functions $\langle V_+^2 \rangle ,\, \langle H_+^2 \rangle ,\, \langle \Delta_+^2 \rangle ,\, \langle H_\times^2 \rangle ,\, \langle \Delta_\times^2 \rangle$ with respect to the standard inflationary results, and it introduces the nondiagonal correlations $\langle V_+ \, H_+ \rangle ,\, \langle V_+ \, \Delta_+ \rangle ,\, \langle H_+ \Delta_+ \rangle ,\, \langle H_\times \Delta_\times \rangle$, which are absent in standard inflation.

As the universe isotropizes, the two modes $\Delta_+$ and $\Delta_\times$ become the two transverse polarizations of the vector field. These modes rapidly decrease after inflation (when the evolution of the vector becomes standard). Therefore, all correlators involving these modes become negligible at late times, and we disregard them in the remainder of this work. The remaining correlators can be written as
in eq. (\ref{corr-aniso}), in terms of the power spectra
\begin{eqnarray}
P_{\cal RR} &=& \frac{1}{2 \pi^2} \, \frac{H^2}{\dot{\phi}^2} \, p^3 \, \left( \Upsilon_s \, \Upsilon_s^\dagger \right)_{11} \nonumber\\
P_{{\cal R} h_+} &=&- \frac{1}{\sqrt{2} \pi^2} \, \frac{H}{\dot{\phi} \, M_p} \, p^3 \, {\rm Re } \left[ \left( \Upsilon_s \, \Upsilon_s^\dagger \right)_{12} \right] \nonumber\\
P_{h_+ h_+} &=& \frac{1}{\pi^2 \, M_p^2} \, p^3 \, \left(  \Upsilon_s \, \Upsilon_s^\dagger \right)_{22} \nonumber\\
P_{h_\times h_\times} &=& \frac{1}{\pi^2 \, M_p^2} \, p^3 \, \left(  \Upsilon_v \, \Upsilon_v^\dagger \right)_{11}
\label{power-spectra}
\end{eqnarray}
To evaluate the power spectra, we impose the initial conditions on $\Upsilon$ as discussed at the end of the previous Subsection. We then evolve $\Upsilon$ through their equations of motion (\ref{eom-upsilon}). We remark that the resulting power spectra are dimensionless, and are written in terms of only physical quantities (namely, they are insensitive to the normalization of the scale factors; this is because only physical quantities appear in the initial conditions and the evolution equations for $\Upsilon$). 

\begin{figure}[h]
\centerline{
\includegraphics[width=0.4\textwidth,angle=-90]{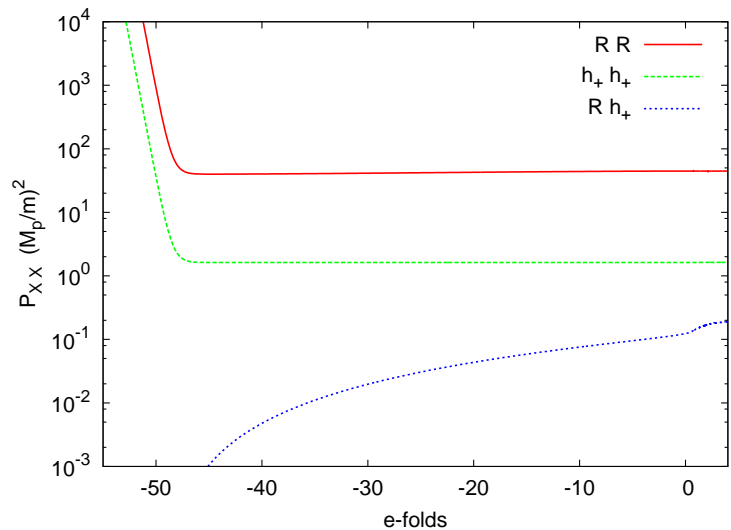}
}
\caption{Time evolution for the power of a specific mode, on a nearly isotropic background ($c-1=10^{-5}$). The number of e-folds $\alpha$ is used as a ``time'' variable, and it is normalized to $0$ at the end of inflation. The mode shown leaves the horizon about $50$ e-folds before the end of inflation. See the main text for details.
}
\label{fig:spe-evol} 
\end{figure}

\begin{figure}[h]
\centerline{
\includegraphics[width=0.4\textwidth,angle=-90]{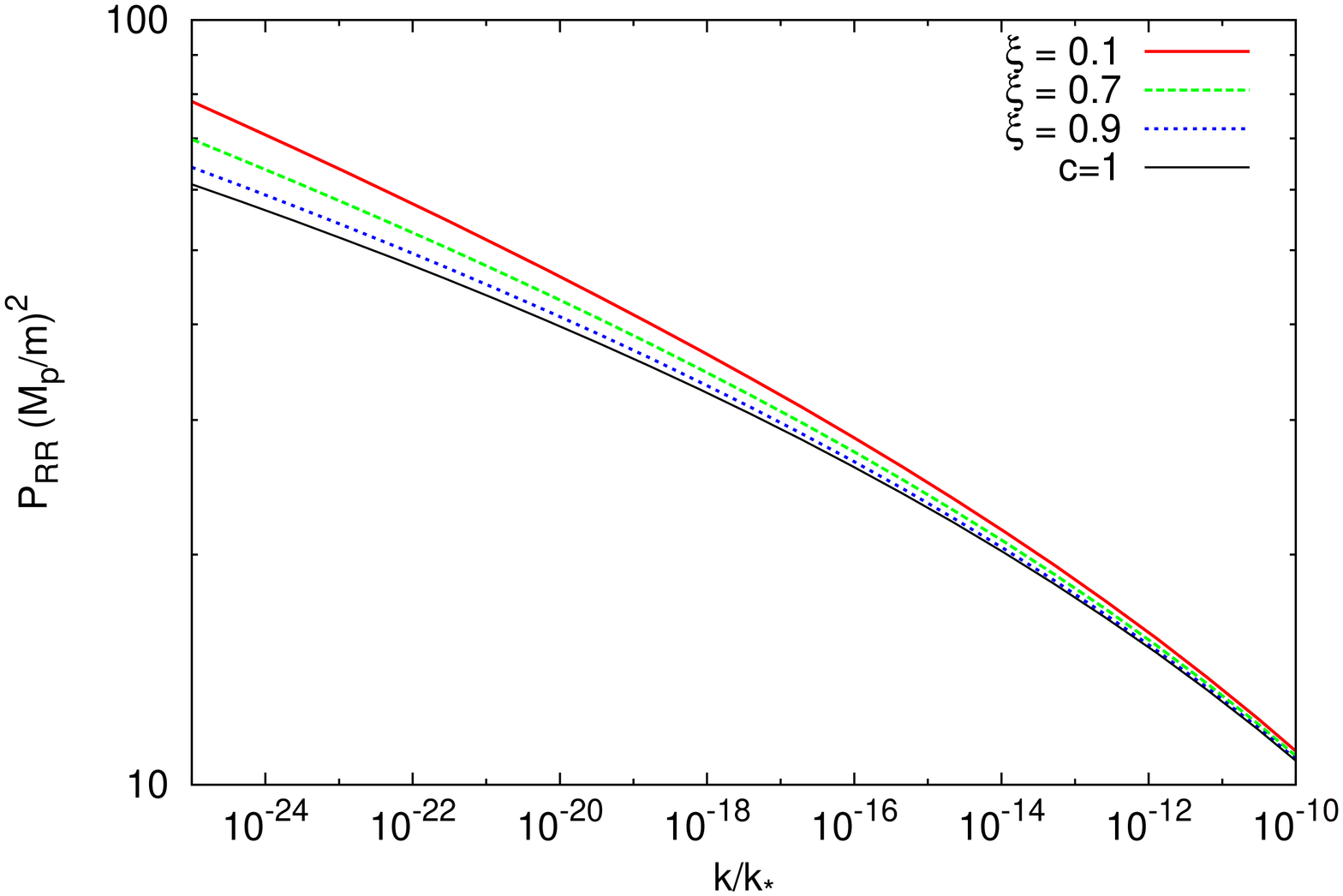}
\includegraphics[width=0.4\textwidth,angle=-90]{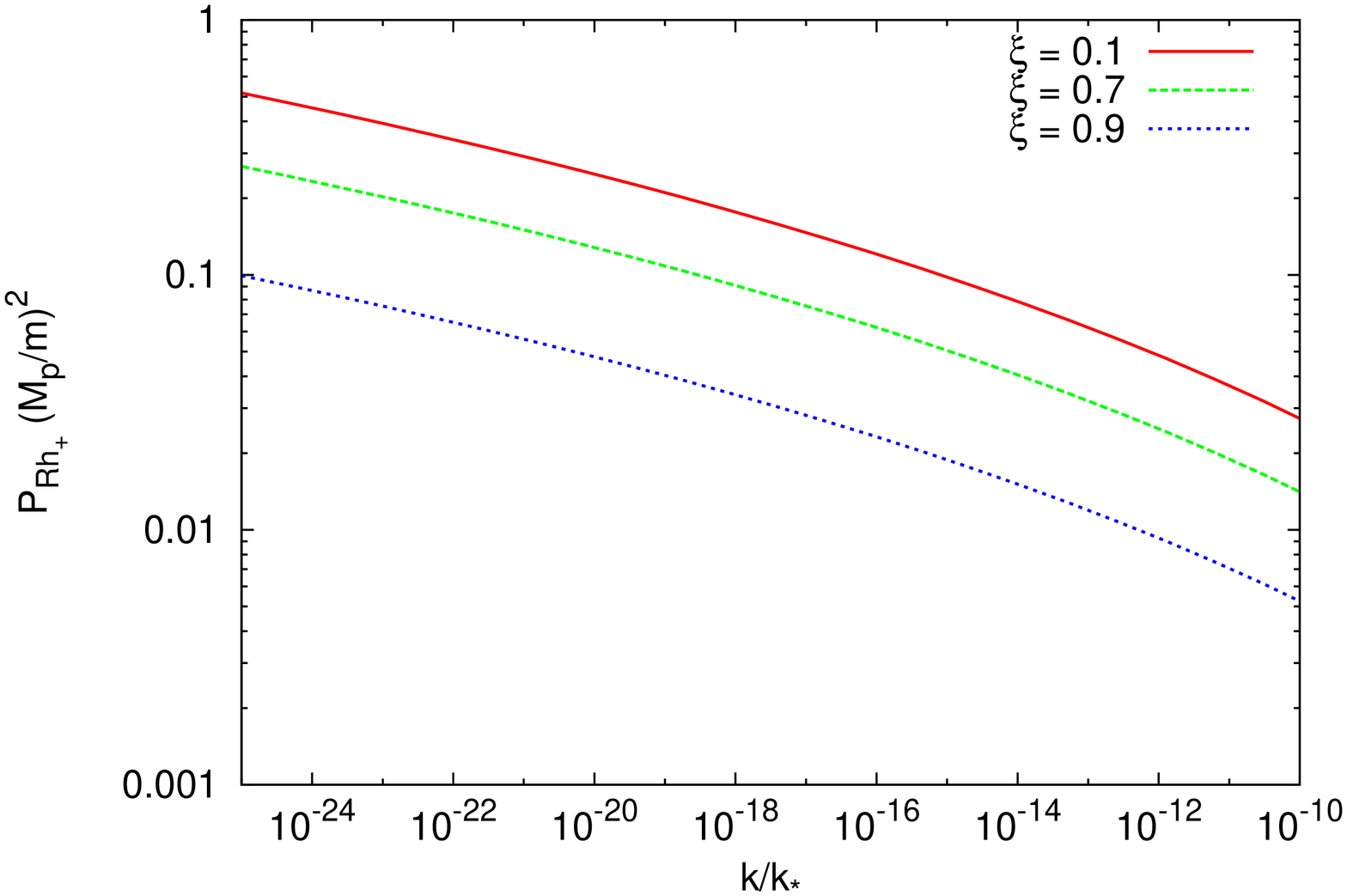}
}
\centerline{
\includegraphics[width=0.4\textwidth,angle=-90]{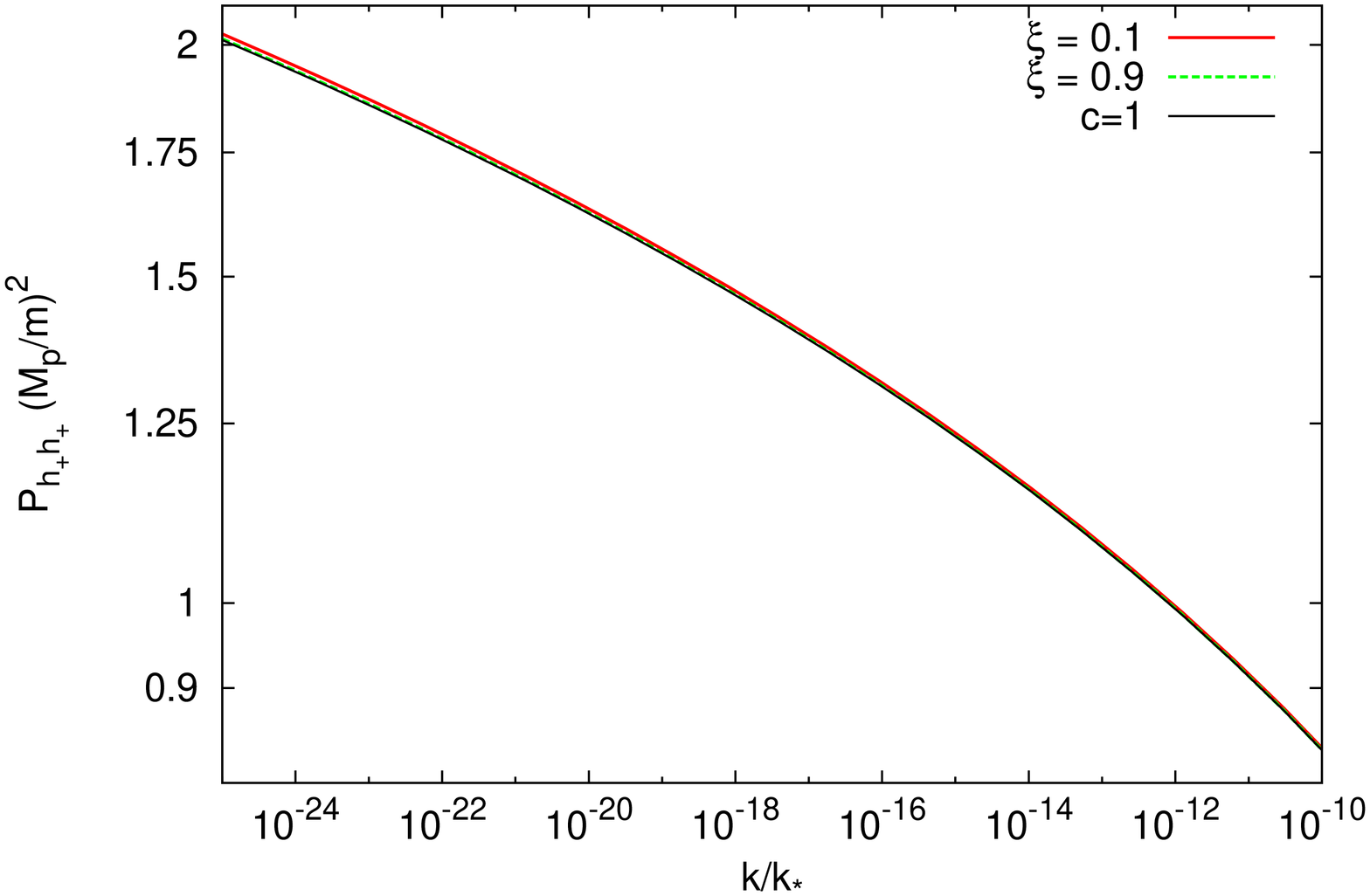}
\includegraphics[width=0.4\textwidth,angle=-90]{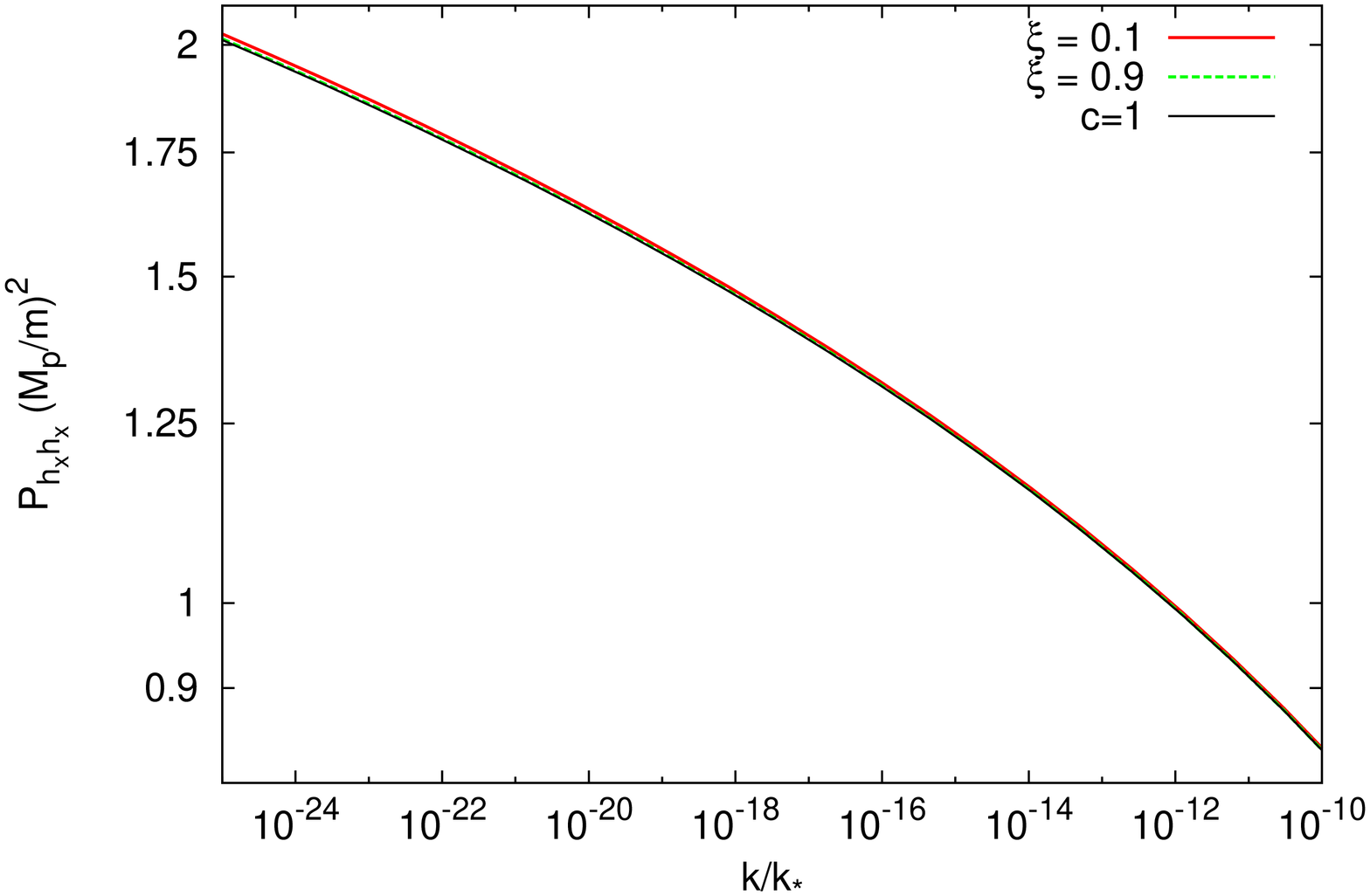}
}
\caption{Scalar-scalar, scalar-tensor, and tensor-tensor power spectra for a nearly isotropic background ($c-1=10^{-5}$). Notice that the scale on the $y$ axis is different for the different panels. For the scalar-scalar, and the 
tensor-tensor spectra, we also show the standard result for comparison (in the present model, the isotropic limit is reached for $c=1$). 
}
\label{fig:spectra} 
\end{figure}

\begin{figure}[h]
\centerline{
\includegraphics[width=0.4\textwidth,angle=-90]{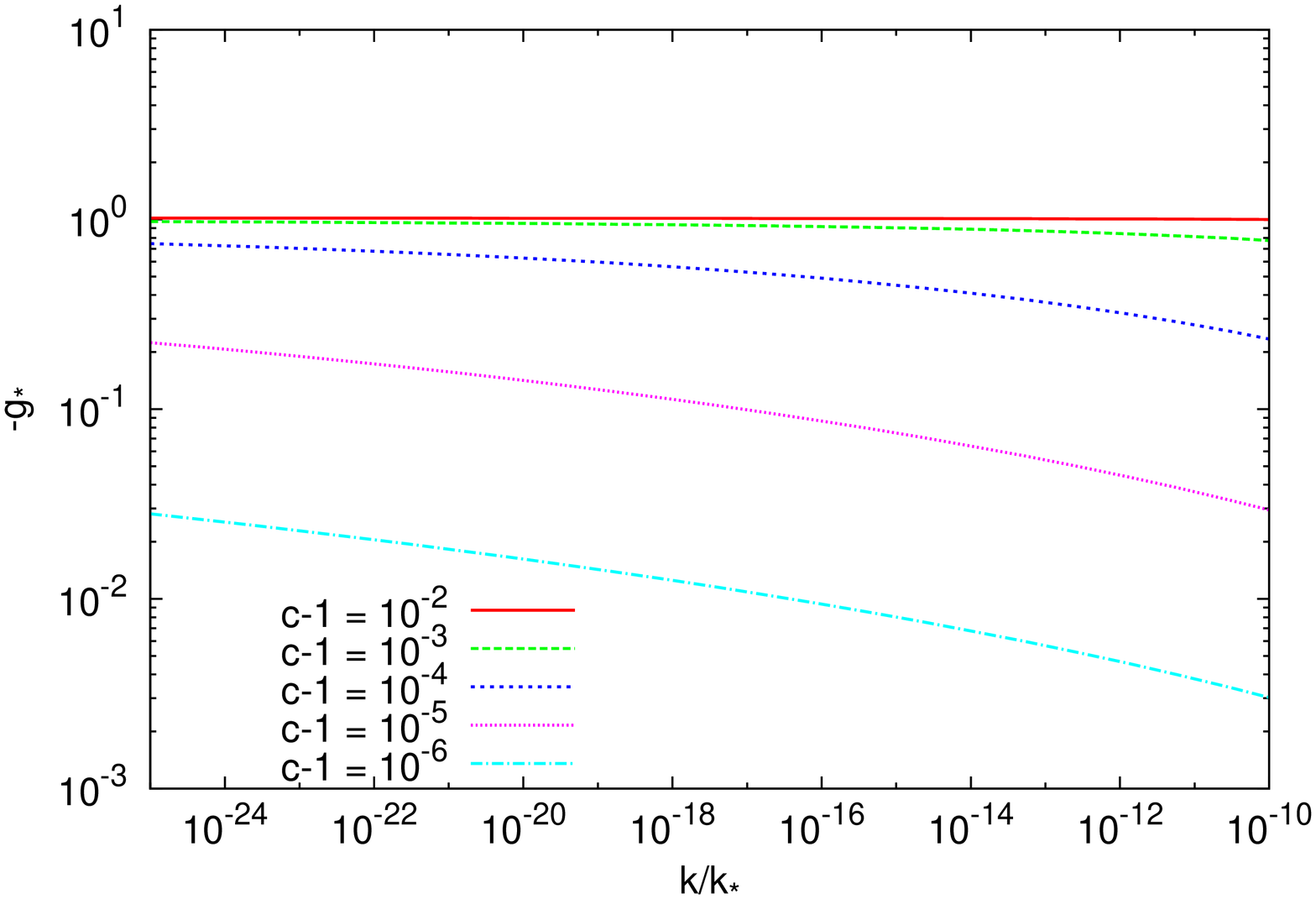}
}
\caption{Angular dependence (expressed through the ACW $g_*$ factor) of the scalar-scalar power spectrum for different choices of $c$. 
}
\label{fig:gstar} 
\end{figure}

Our results are shown in Figures \ref{fig:spe-evol}, \ref{fig:spectra}, and \ref{fig:gstar}. Figure \ref{fig:spe-evol} shows the evolution of the power for the specific value of $c - 1 = 10^{-5}$ in the kinetic function (\ref{Vf}), and for a specific mode: we denote by $k_*$ the magnitude of the momentum of a mode which barely exits the horizon at the end of inflation; we choose a mode with momentum $k = 10^{-20} \, k*$, and we choose the cosine of the angle between the momentum and the privileged direction to be $\xi = 1/2$ (such values have no particular meaning, and are just chosen for illustrative purposes; different values of $k$ and $\xi$ lead to the same qualitative behavior); this mode is initially deeply inside the horizon, and it leaves the horizon about $50$ e-folds before the end of inflation.~\footnote{Since this choice of $c \simeq 1$ corresponds to a very small anisotropy, we define the horizon, and the number of e-folds, only through the evolution of the ``average'' scale factor ${\rm e}^\alpha$, as in FRW cosmology.} 
The power in the scalar-scalar and tensor-tensor correlations behaves analogously to the isotropic case.
The power in the scalar-tensor cross correlation is instead very small initially, and slightly increases during and immediately after inflation. The power becomes constant (at the value seen in the latest time shown in the Figure) as the universe isotropizes after inflation.

In Figure \ref{fig:spectra} we show the power spectra for the specific value of $c - 1 = 10^{-5}$ in the kinetic function (\ref{Vf}). The smallest momentum shown corresponds to modes that exited the horizon about $60$ e-folds before the end of inflation. For the scalar-scalar and tensor-tensor case, the standard result is also shown for comparison. The scalar-scalar power spectrum (bottom left panel) is slightly greater than in the isotropic case; this can be compensated by decreasing the scalar field mass (for this reason, the ratio $m/M_p$ has been kept as a free parameter in the Figure). The angular dependence of $P_{\cal RR}$ is of ${\rm O } \left( 10^{-1} \right)$ at the largest scales, while it slowly decreases at greater scales. The tensor-tensor power spectra (bottom panels) are closer to the standard result, and
they exhibit a much milder angular dependence (we found $g_*$ for the scalar spectra is suppressed with respect to $g_*$ in the tensor spectrum by approximately the ratio between the power of the tensor and  the scalar spectra). Moreover, the results for the two polarizations are nearly identical. The scalar-tensor cross correlation (upper right panel) shows a stronger angular dependence, but it is smaller than the other two spectra.

In Figure \ref{fig:gstar} we show the angular dependence $g_*$ (defined  in equation (\ref{acw})) for the scalar-scalar power spectrum, for different values of $c$ in eq. (\ref{Vf}). We remark that all values of $c$ provide a negative $g_*$ (in Figure \ref{fig:gstar} we actually show $\vert g_* \vert = -  g_*$ in logarithmic scale). The value of $g_*$ shown in the plots is obtained by comparing, for each  value of $k = \vert {\bf k} \vert$, the power at $\xi = 0.1$ and at  $\xi = 0.9$. We have however verified that the ACW parametrization (\ref{acw}) is very accurate, in the sense that, once $g_*$ and $P \left( k \right)$ are obtained from the results at  $\xi = 0.1$ and $\xi = 0.9$, also other values of $\xi$ are fitted very well by (\ref{acw}). 

The significance of the results summarized in these Figures is discussed in the next concluding Section.

\section{Discussion}

\label{sec:conclusion}

In this work we compute the precise phenomenological signatures for the 
model of \cite{Watanabe:2009ct}. This model admits an anisotropic inflationary background evolution, supported by the combined presence of a scalar and a vector field. This solution is (mathematically) continuously connected to an isotropic solution, in the limit in which the energy density associated to the vector is sent to zero. Moreover, the model is free of the ghost instabilities that plague other models with vector fields during inflation, due to the U(1) invariance of its action under a shift of the vector. Therefore, it offers a complete, and stable counterexample to Wald's no hair theorem on the isotropization of Bianchi spaces \cite{wald}.

We studied the simplest realization of the idea of \cite{Watanabe:2009ct}, in which a single vector is present, and the potential of the scalar is taken to be that of massive chaotic inflation. We found that the scalar-scalar correlation function exhibits an angular dependence which is however of the wrong sign to account for the breaking of rotational invariance seen in the data: the model gives a negative value for $g_*$, while the analysis of \cite{Groeneboom:2009cb} indicates that $g_* = 0.29 \pm 0.031$ in the WMAP W-band. We also found that the amount of anisotropy in the spectrum (the order of magnitude of $\vert g_* \vert$) is not of the same order of magnitude as  the amount of anisotropy in the expansion (the order of magnitude of $\Delta H/ H \sim \dot{\sigma} / \dot{\alpha}$, where $H$ is the average expansion rate, and $\Delta H$ the difference between the expansion rate in the different coordinates). A $\vert g_* \vert = {\rm O } \left( 10^{-1} \right)$ is obtained when the anisotropy in the expansion is of ${\rm O } \left( 10^{-7}  - 10^{-6} \right)$ during inflation. It is also worth noting that the anisotropy in the tensor-tensor spectrum is much milder than that in the scalar-scalar spectrum (we find that the suppression is approximately proportional to the ratio between the two power spectra).

A distinctive feature of the model (and, of anisotropic spaces in general) is a nonvanishing scalar-tensor correlation, which, if sufficiently high, may be detected through temperature-B mode correlation in the CMB. A naive estimate actually suggests that the scalar-tensor correlation could be higher than the tensor-tensor one. Indeed the amplitude of a tensor mode is approximately multiplied by $2 \, \sqrt{\epsilon}$ with respect to that of a scalar mode, where $\epsilon \equiv - \dot{H}/H^2$ is a slow roll parameter.~\footnote{Since we are considering a small anisotropy, we can use the FRW computation in this estimate; then, comparing the two equations in (\ref{eq-iso}) we see that $\vert V_+ \vert \simeq \vert H_{+,\times} \vert$. This gives $\vert h_{+,\times} / {\cal R} \vert \simeq \sqrt{2} \dot{\phi} / \left( M_p \, H \right) \simeq 2 \epsilon$.} The tensor-tensor correlator is suppressed by one more power of this factor than the tensor-scalar correlator. However, the scalar-tensor correlator vanishes for an isotropic space, and hence must be suppressed by some factor related to the asymmetry. If one naively assumes that this factor is $\vert g_* \vert$, one would get the prediction that $P_{{\cal R}h} / P_{hh} \simeq \vert g_* \vert / \left( 2 \sqrt{\epsilon} \right)$. One would then get $P_{{\cal R}h} / P_{hh} \simeq 6 \,\vert g_*\vert$ for chaotic inflation, and a greater value for other inflationary models, characterized by a smaller value of $\epsilon$. Our explicit results show that this estimate is reasonable, but not exact. For the case of $c-1=10^{-5}$ shown in Figure \ref{fig:spectra} (giving $g_* \simeq -0.23$), the estimate gives $P_{{\cal R}h} / P_{hh} \simeq 1.4$, while the actual spectra give $P_{{\cal R}h} / P_{hh} \simeq 0.05-0.25$, depending on the orientation of the mode (the value of $\xi$).

To conclude, while the simplest realization of \cite{Watanabe:2009ct} cannot explain the breaking of rotational invariance seen in the data, and, most likely, it does not give rise to an interesting scalar-tensor correlation, this model is, to our knowledge, the first complete and stable model of anisotropic inflation for which the phenomenological predictions strictly follow from the action (and not from arbitrary initial conditions), and have been computed. As the observed breaking of rotational invariance awaits for a confirmation, or a refutation, from Plank, our work provides the tools for studying different models, to see whether they can reproduce the WMAP feature, and perhaps lead to new predictions.

\vspace{1cm}
{\bf Note added: } After the completion of the analysis reported here, and as we were finalizing the preparation of the present manuscript for submission, ref. \cite{Dulaney:2010sq} appeared on the preprint archive, which also computed some of the correlation functions computed here. Why we obtained the expressions for the correlators analytically, and we then evaluated them numerically, 
ref. \cite{Dulaney:2010sq} performs a fully analytical computation, treating the anisotropy as a small perturbation. The relation between $g_*$ and $\dot{\sigma} / \dot{\alpha}$ found in \cite{Dulaney:2010sq} and in the present work are in very good agreement with each other. The two studies also agree on the fact that the tensor-tensor correlator shows a smaller angular dependence than the scalar-scalar one. The scalar-tensor correlator has not been computed in \cite{Dulaney:2010sq}.

\begin{acknowledgments}

The work of A.E.G. and of M.P. was partially supported by the DOE grant DE-FG02-94ER-40823.
The work of B.H. was supported by the Graduate School at the University of Minnesota under the Doctoral Dissertation Fellowship.

\end{acknowledgments}

\appendix

\section{Explicit quadratic action of the perturbations} \label{appendix-explicit}

We provide here the explicit expressions for the matrices entering in the action and the evolution equations of the canonical modes, eqs. (\ref{action-formal}) and (\ref{eom-formal}).

For the $2$d scalar modes we have

\begin{eqnarray}
&&
K_s = \left( \begin{array}{ccc}
0 & 0 & K_{13,s} \\
0 & 0 & K_{23,s} \\
- K_{13,s} & - K_{23,s} & 0
\end{array} \right)
\;\;\;,\;\;\;
\Omega_s^2 = \left( \begin{array}{ccc}
\Omega_{11,s}^2 & \Omega_{12,s}^2 & \Omega_{13,s}^2 \\
\Omega_{12,s}^2 & \Omega_{22,s}^2 & \Omega_{23,s}^2 \\
\Omega_{13,s}^2 & \Omega_{23,s}^2 & \Omega_{33,s}^2 \\
\end{array} \right)
\end{eqnarray}
where
\begin{eqnarray}
K_{13,s} = - \frac{{\tilde p}_A \, p_T \, f' \left( \phi \right)}{p \, f \left( \phi \right)^2}
\;\;,\;\;
K_{23,s} = - \frac{{\tilde p}_A \, p_T }{\sqrt{2} \, M_p \, p \, f \left( \phi \right)} \nonumber
\end{eqnarray}
\begin{eqnarray}
\Omega_{11,s}^2 &=& p^2 - \frac{9}{4}\, \dot\alpha^2 + \frac{15\,
\dot\phi^2}{4 M_p^2}  +
\frac{9}{2}\, \dot\sigma^2 + \frac{2\, \dot\phi\,
V'(\phi)}{M_p^2\, \dot\alpha} + V''(\phi) -2\, \frac{p^4}{{\cal
D}^2}\, \frac{\dot\phi^4}{M_p^4} 
\nonumber\\
&& + \frac{2\, \left( p_T^2 - 2 p_L^2 \right)\, V'(\phi)}{{\cal
D}}\, \frac{\dot\phi}{M_p^2\, \dot\alpha}\, \dot\sigma - 3\,
\left( 4\, \frac{\left( 2 p_L^2 - p_T^2 \right)\, p^2}{{\cal
D}^2}\, \dot\alpha + \frac{8 p_L^4 - 8 p_L^2\, p_T^2 + 5
p_T^4}{{\cal D}^2}\, \dot\sigma \right)\,
\frac{\dot\phi^2}{M_p^2}\, \dot\sigma \nonumber\\
&& + \frac{\tilde{p}_A^2}{2 M_p^2\, f(\phi)^2}\, \Big[ 1 + \frac{2
M_p^2\, \left( 3 p_L^2 - p_T^2 \right)\, f'(\phi)^2}{p^2\,
f(\phi)^2} - 4\, \frac{p_T^2\, p^2}{{\cal D}^2}\,
\frac{\dot\phi^2}{M_p^2} - 8\, \frac{p_L^2}{{\cal D}}\,
\frac{f'(\phi)}{f(\phi)}\, \dot\phi - 2 M_p^2\,
\frac{f''(\phi)}{f(\phi)} \Big] \nonumber
\end{eqnarray}
\begin{eqnarray}
{\Omega}_{12,s}^2 &=& -\frac{3\, \sqrt{2}\, p_T^2\, p^2\,
\dot\sigma}{{\cal D}^2}\, \left[ \frac{\dot\phi^3}{M_p^3} 
- 6\, \frac{\dot\phi}{M_p}\,
\left( \dot\alpha + \dot\sigma \right)\, \left( \dot\alpha +
\frac{p_L^2 - p_T^2}{p^2}\, \dot\sigma \right) -
\frac{V'(\phi)}{M_p}\, \left( 2 \dot\alpha + \frac{2 p_L^2 -
p_T^2}{p^2}\, \dot\sigma \right) \right] \nonumber\\
&& - \frac{\sqrt{2}\, \tilde{p}_A^2\, p_T^2}{M_p^2\, f(\phi)^2\,
{\cal D}^2}\, \Bigg[ 3 p_T^2\, \frac{\dot\phi}{M_p}\, \dot\sigma +
M_p\, \frac{f'(\phi)}{f(\phi)}\, \Big( 4\, p^2\, \dot\alpha^2 +
2\, \left( 7 p_L^2 - 2 p_T^2 \right)\,
\dot\alpha\, \dot\sigma \nonumber\\
&& \qquad\qquad\qquad\qquad\qquad\qquad\qquad  + \frac{\left( 2
p_L^2 - p_T^2\right)\, \left( 5 p_L^2 - p_T^2\right)}{p^2}\,
\dot\sigma^2 \Big) \Bigg] \nonumber
\end{eqnarray}
\begin{eqnarray}
{\Omega}_{13,s}^2 &=& \frac{ \tilde{p}_A\, p_T}{p\, f(\phi)}\,
\Bigg\{ - \frac{2\, \tilde{p}_A^2\, p^2}{M_p^2\, f(\phi)^2\, {\cal
D}^2}\, \left[ \frac{\dot\phi}{M_p^2}\, p_T^2 +
\frac{f'(\phi)}{f(\phi)}\, p_L^2\, \left( 2 \dot\alpha + \frac{2
p_L^2 - p_T^2}{p^2}\, \dot\sigma \right) \right]
\nonumber\\
&& -2\, \frac{p^4}{{\cal D}^2\, M_p}\, \left[
\frac{\dot\phi^3}{M_p^3}  -
6\, \frac{\dot\phi}{M_p}\, \left( \dot\alpha + \dot\sigma
\right)\, \left( \dot\alpha + \frac{p_L^2-
p_T^2}{p^2}\, \dot\sigma \right) \right] \nonumber\\
&& + 2\, \frac{p^4}{{\cal D}^2\, M_p}\, \frac{V'(\phi)}{M_p}\,
\left( 2\, \dot\alpha + \frac{2 p_L^2 - p_T^2}{p^2}\, \dot\sigma
\right) - \frac{f'(\phi)}{f(\phi)}\, \left( \dot\alpha + \frac{7
p_L^2 - 2 p_T^2}{p^2}\, \dot\sigma \right) + \dot\phi\,
\frac{f''(\phi)}{f(\phi)} \Bigg\} \nonumber
\end{eqnarray}
\begin{eqnarray}
\Omega_{22,s}^2 &=& p^2 - \frac{1}{4}\, \dot\alpha^2 + \frac{3\,
\dot\phi^2 }{4 M_p^2} - 8\, \frac{p^4}{{\cal
D}^2}\, \dot\alpha^4 - \frac{9 p_T^4\, \dot\sigma^2}{{\cal D}^2}\,
\frac{\dot\phi^2 }{M_p^2} - 8\, \frac{\left( 2 p_L^2
- p_T^2 \right)\, p^2}{{\cal D}^2}\, \dot\alpha^3\,
\dot\sigma \nonumber\\
&& + 2\, \frac{5 p_L^4 + 58\, p_L^2\, p_T^2 + 35\, p_T^4}{{\cal
D}^2}\, \dot\alpha^2\, \dot\sigma^2 + 18\, \frac{2 p_L^4 + 9\,
p_L^2\, p_T^2 - p_T^4}{{\cal D}^2}\, \dot\alpha\, \dot\sigma^3 
\nonumber\\
&&
+
\frac{9}{2}\, \frac{4 p_L^4 + 12\, p_L^2\, p_T^2 - 11 p_T^4}{{\cal
D}^2}\, \dot\sigma^4 
+ \frac{\tilde{p}_A^2}{2 M_p^2\, f(\phi)^2}\, \frac{p_L^2 -
p_T^2}{p^2} - \frac{9\, \tilde{p}_A^2}{M_p^2\, f(\phi)^2}\,
\frac{p_T^6}{p^2\, {\cal D}^2}\, \dot\sigma^2 \nonumber
\end{eqnarray}
\begin{eqnarray}
\Omega_{23,s}^2 &=& \frac{ \tilde{p}_A\, p_T}{\sqrt{2}\, M_p\,
p\, f(\phi)}\, \Bigg\{ -\frac{6 \tilde{p}_A^2}{M_p^2\,
f(\phi)^2}\, \frac{p_T^4}{{\cal D}^2}\, \dot\sigma +
\frac{f'(\phi)}{f(\phi)}\, \dot\phi - 4\, \frac{p^4}{{\cal
D}^2}\, \dot\alpha^3 \nonumber\\
&& \qquad\qquad\qquad\qquad - 6\, \frac{p_T^2\, p^2}{{\cal D}^2}\,
\left( \frac{\dot\phi^2 }{M_p^2} - 2\, \frac{p_L^2 +
4 p_T^2}{p_T^2}\, \dot\alpha^2 \right)\, \dot\sigma + 9\, \frac{4
p_L^4 + 16\, p_L^2\, p_T^2 - p_T^4}{{\cal D}^2}\,
\dot\alpha\, \dot\sigma^2 \nonumber\\
&& \qquad\qquad\qquad\qquad + \frac{20 p_L^6 + 96 p_L^4\, p_T^2 -
39 p_L^2\, p_T^4 - 34 p_T^6}{p^2\, {\cal D}^2}\, \dot\sigma^3
\Bigg\} \nonumber
\end{eqnarray}
\begin{eqnarray}
\Omega_{33,s}^2 &=& p^2 + \frac{\dot\phi^2 }{4 M_p^2}
- \frac{1}{4}\, \dot\alpha^2 - \frac{p_L^2 - 2 p_T^2}{2 p^2}\,
\left( 4\, \dot\alpha - \frac{p_L^4 + 50 p_L^2\, p_T^2 - 5
p_T^4}{p^2\, \left( p_L^2 -2 p_T^2\right)}\, \dot\sigma \right)\,
\dot\sigma \nonumber\\
&& + \left[ V'(\phi) + 2\, \dot\phi\, \dot\alpha + 2\, \frac{p_L^2
- 2 p_T^2}{p^2}\, \dot\phi\, \dot\sigma \right]\,
\frac{f'(\phi)}{f(\phi)} - \dot\phi^2\,
\frac{f''(\phi)}{f(\phi)} \nonumber\\
&& - \frac{\tilde{p}_A^2\, p^2\, p_T^2}{2 M_p^2\, f(\phi)^2\,
{\cal D}^2}\, \Bigg[ \frac{4\, \tilde{p}_A^2}{M_p^2\, f(\phi)^2}\,
\frac{p_T^2}{p^2} + \frac{2 M_p^2\, {\cal D}^2}{p^2\, p_T^2}\,
\frac{f'(\phi)^2}{f(\phi)^2} + \frac{4\,  \dot\phi^2 
}{M_p^2} - 20\,
\frac{p^2}{p_T^2}\, \dot\alpha^2 \nonumber\\
&& - 4\, \frac{10 p_L^4 + 17\, p_L^2\, p_T^2 + p_T^4}{p^2\,
p_T^2}\, \dot\alpha\, \dot\sigma - \frac{\left( 2 p_L^2 + 5 p_T^2
\right)\, \left( 10 p_L^4 - p_L^2\, p_T^2 - 5 p_T^4 \right)}{p^4\,
p_T^2}\, \dot\sigma^2 \Bigg] \nonumber\\\label{Omega2}
\end{eqnarray}
and where we have defined
\begin{equation}
{\cal D} \equiv 2 p^2\, \dot\alpha + \left( 2 p_L^2 - p_T^2
\right)\, \dot\sigma \label{calD}
\end{equation}

For the $2$d vector modes we have instead
\begin{eqnarray}
&&K_v = \left( \begin{array}{cc} 
0 & K_{12,v} \\ - K_{12,v} & 0 
\end{array} \right) \;\;\;,\;\;\;
\Omega_v^2 = \left( \begin{array}{cc} 
\Omega_{11,v}^2 & \Omega_{12,v}^2 \\
\Omega_{12,v}^2 & \Omega_{22,v}^2
\end{array} \right) \nonumber\\
\end{eqnarray}
where
\begin{eqnarray}
&&K_{12,v} = - \frac{{\tilde p}_A \, \vert p_T \vert}{\sqrt{2} \, M_p \, f \left( \phi \right) \, p }
\nonumber\\
&&\Omega_{11,v}^2 = p^2 - \frac{9}{4} \dot{\alpha}^2 -  \dot{\sigma}^2 \left[  \frac{9}{2} - 36 \, \frac{p_L^2}{p^2}+ 27 \, \frac{p_L^4}{p^4} \right] + \frac{3 \dot{\phi}^2}{4 M_p^2} - \frac{{\tilde p}_A^2}{2 M_p^2 f \left( \phi \right)^2} \, \left[ 1 - 2 \, \frac{p_L^2}{p^2} \right] \nonumber\\
&&\Omega_{22,v}^2 = p^2 - \frac{\dot{\alpha}^2}{4} - 2 \, \dot{\alpha} \, \dot{\sigma} + \frac{\dot{\sigma}^2}{2} 
+ \frac{\dot{\phi}^2}{4 \, M_p^2} + \frac{{\tilde p}_A^2}{2 M_p^2 f \left( \phi \right)^2} 
\left[ 1 + 4 \, \frac{p_L^2}{p^2} \right]
\nonumber\\
&&\quad\quad\;\; + \left[ V' \left(\phi \right) + 2 \left( \dot{\alpha} + \dot{\sigma} \right) \dot{\phi} \right] \frac{f' \left( \phi \right)}{f \left( \phi \right)} - {\tilde p}_A^2 \, \frac{f'\left(\phi\right)^2}{f\left(\phi\right)^4}
- \dot{\phi}^2 \, \frac{f''\left(\phi\right)}{f\left(\phi\right)} \nonumber\\
&&\Omega_{12,v}^2 = \frac{{\tilde p}_A \, \vert p_T \vert}{\sqrt{2} \, M_p \, f \left( \phi \right) p} \, \left[
-\dot{\alpha} + \left( 9 \frac{p_L^2}{p^2} - 4 \right) \dot{\sigma} 
+ \frac{\dot{\phi} \, f' \left( \phi \right)}{f \left( \phi \right)} \right]
\end{eqnarray} 

The physical momenta entering in these expressions are related to the comoving ones given in the main text by
\begin{equation}
p_L \equiv \frac{k_L}{a \left( t \right)} \;\;\;,\;\;\;
p_T \equiv \frac{k_T}{b \left( t \right)} \;\;\;,\;\;\;
p \equiv \sqrt{p_L^2+p_T^2}
\end{equation}

\section{Gauge invariant perturbations in terms of perturbations in our gauge, and late time isotropic limit} \label{appendix-iso}

In this Appendix we discuss the late time interpretation of the perturbations in the gauge chosen in the main text. This interpretation is done when the background has become isotropic, and the vev of the vector has gone to zero. In this case, the Fourier coefficients of our metric perturbations read
\begin{equation}
\delta g_{\mu \nu} \left( k \right) = \left( \begin{array}{cccc}
- 2 \Phi & i \, a \, k_L \, \chi & 
i a \left( k_{T2} \, B + k_{T3} \, B_v \right) &
i a \left( k_{T3} \, B - k_{T2} \, B_v \right) \\
& - 2 a^2 \Psi & 
- a^2 \, k_L \, k_{T3} \, {\tilde B}_v & 
 a^2 \, k_L \, k_{T2} \, {\tilde B}_v \\
& & 0 & 0 \\
& & & 0 
\end{array} \right) \;\;\;,\;\;\;
\label{ourmodes}
\end{equation}
as can be seen by imposing the gauge (\ref{ourgauge}) on the parametrization given by (\ref{perts-def1})
and (\ref{2dv-transv}). We recall that $\left( k_L ,\, k_{T2} ,\, k_{T3} \right)$ denotes the comoving momentum of the mode we are studying.~\footnote{In the main text, the suffices $L$ and $T$ refer to ``longitudinal'' or ``transverse'' with respect to direction $x$, which was the anisotropic one during inflation.} We perform a rotation to a coordinate system for which the momentum is along the third direction, $k_\mu \rightarrow {\tilde k}_\mu = R_\mu^\nu \, k_\nu =  \left( 0 ,\, 0 ,\, \sqrt{k_L^2 + k_{T2}^2+k_{T3}^2} \right) \,$. The explicit form of the rotation matrix is
\begin{equation}
R^\mu_\nu = 
\left( \begin{array}{cccc}
1 & 0 & 0 & 0 \\
0 & - \frac{k_T}{k} & \frac{k_L \, k_{T2}}{k_T \, k} & \frac{k_L \, k_{T3}}{k_T \, k} \\ 
0 & 0 & \frac{k_{T3}}{k_T} & - \frac{k_{T3}}{k_T} \\
0 & \frac{k_L}{k} & \frac{k_{T2}}{k} & \frac{k_{T3}}{k}
\end{array} \right) 
\end{equation}
where $k_T = \sqrt{k_{T2}^2 + k_{T3}^2}$, and the metric transforms as
\begin{equation}
g_{\mu \nu} \rightarrow {\widetilde g}_{\mu \nu} = R_\mu^\alpha \, 
g_{\alpha \beta} \, R_\nu^\beta\,
\label{rotation}
\end{equation}

Although the gauge in which eq. (\ref{ourmodes}) appears is nonstandard, we can combine the metric perturbations  (\ref{ourmodes}) into the gauge invariant expressions that are commonly used. These gauge invariant combinations are usually expressed starting from the most general metric perturbations, classified as scalar, vector, or tensor, with respect to $3$d spatial rotations:
\begin{equation}
\delta g_{00} = - 2 \Phi_* \;\;\;,\;\;\;
\delta g_{0i} = 2 a \left( B_{*i} + \partial_i B_* \right)  \;\;\;,\;\;\;
\delta g_{ij} = a^2 \left[  - 2 \Psi_* \, \delta_{ij} + 2 \, E_{*,ij} + E_{*(i,j)} + h_{*ij} \right] 
\label{general-perts}
\end{equation}
where $i=1,2,3$. The ($3$d) vectors $B_{*i}$ and $E_{*i}$ are transverse, while the tensor mode $h_{*ij}$ is transverse and traceless. The remaining modes are scalar, and are coupled also to the perturbation of the scalar field $\delta \phi_*$.  Out of these modes, we are interested in the the gauge invariant scalar combination \cite{Mukhanov:1990me,Riotto:2002yw}
\begin{equation}
{\cal R} \equiv \Psi_* + \frac{H}{\dot{\phi}} \delta \phi_*
\end{equation}
and in the two (gauge invariant) tensor mode polarizations $h_+$ and $h_\times$ encoded in $h_{*ij}$ (the vector modes disappear once the universe becomes isotropic). 

Expression (\ref{general-perts}) gives the metric perturbations before any gauge is chosen. By equating them with our expressions (\ref{ourmodes}) we find how our modes can be decomposed into $3$d scalar, vector, and tensor perturbations. We can then use the resulting expressions to write ${\cal R}$, $h_+$ and $h_\times$ in terms of our modes. To make this identification, we should spell out explicitly how the components of the $3$d vector and tensor modes enter in (\ref{general-perts}), accounting for their transversality, and traceless properties. We do so in the ${\tilde x}^\mu$ coordinate system, for which the momentum of the mode is ${\tilde k}_\mu = \left( 0 ,\, 0 ,\, \sqrt{k_L^2 + k_{T2}^2+k_{T3}^2} \right) \,$. In this system, eqs. (\ref{general-perts}) give
\begin{equation}
{\widetilde \delta g}_{\mu \nu} \left( k \right) = \left( \begin{array}{cccc}
- 2 \Phi_* & a \, B_{*1} & a \, B_{*2} & a \, i \, k \, B_* \\
& a^2 \left( - 2 \Psi_* + h_+ \right) & a^2 h_\times & a^2 \, i \, k \, E_{*1} \\
& & a^2 \left( - 2 \Psi_* - h_+ \right) & a^2 \, i \, k \, E_{*2} \\
& & & a^2 \left( - 2 \Psi_* - 2 k^2 E_* \right)
\end{array} \right)
\end{equation}
The entries of this metric can be now identified with those of our metric (\ref{ourmodes}), transformed according to (\ref{rotation}). We obtain
\begin{equation}
{\cal R} = \frac{k_T^2}{2 \, k^2} \Psi + \frac{H}{\dot{\phi}} \delta \phi \;\;\;,\;\;\;
h_+ = - \frac{k_T^2}{k^2} \, \Psi \;\;\;,\;\;\;
h_\times = \frac{k_L \, k_T^2}{k} \, {\tilde B}_v
\end{equation}

Finally, we rewrite our three variables $\Psi ,\, \delta \phi ,$ and ${\tilde B}_v$ in terms of the canonically normalized modes introduced in (\ref{2ds-can}) and (\ref{2dv-can}) (see also eq. (\ref{2dv-transv})). This leads to 
\begin{equation}
R = \frac{H}{a^{3/2} \, \dot{\phi}} \, V_+ \;\;\;,\;\;\;
h_+ = - \frac{\sqrt{2}}{a^{3/2} \, M_p} \, H_+ \;\;\;,\;\;\;
h_\times = \frac{i \, \sqrt{2}}{a^{3/2} \, M_p} \, H_\times
\label{iso-gaugeinv-app}
\end{equation}

As a check, we can verify that the our evolution equations reduce to the standard ones in the limit of isotropic background. Using ${\tilde p}_A = \dot{\sigma} = 0$ in the explicit expressions given in Appendix \ref{appendix-explicit}, we find that $K_{s,v} = 0$ and $\Omega_{s,v}^2$ are diagonal in this limit. Therefore, all the canonical modes are decoupled. One can then show that, in this limit the equations (\ref{eom-formal}) give
\begin{eqnarray}
&&\left( \frac{V_+}{\sqrt{a}} \right)'' + \left[ k^2 - \frac{z''}{z} \right] \left( \frac{V_+}{\sqrt{a}} \right) = 0 \;\;\;,\;\;\;
z \equiv \frac{a^2\,\dot{\phi}}{\dot{a}} \nonumber\\
&&\left( \frac{H_{+,\times}}{\sqrt{a}} \right)'' + \left[ k^2 - \frac{a''}{a} \right] \left( \frac{H_{+,\times}}{\sqrt{a}} \right) = 0 
\label{eq-iso}
\end{eqnarray}
where prime denotes differentiation with respect to conformal time $\eta$, related to the physical time $t$ by $d \eta = d t / a$. From these expressions, and from eqs. (\ref{iso-gaugeinv-app}) we find
\begin{equation}
v'' + \left( k^2 - \frac{z''}{z} \right) v = 0 \;\;\;,\;\;\;
\left( a \, h_{+,\times} \right)'' + \left( k^2 - \frac{a''}{a} \right)
\left( a \, h_{+,\times} \right) = 0
\end{equation}
where $v \equiv z \, {\cal R}$. These equations are the standard ones \cite{Mukhanov:1990me}, confirming that our formalism reduces to the standard one in the limit of isotropic background.


\begin{thebibliography}{0}



\bibitem{Hinshaw:2008kr}
  G.~Hinshaw {\it et al.}  [WMAP Collaboration],
  Astrophys.\ J.\ Suppl.\  {\bf 180}, 225 (2009)
  [arXiv:0803.0732 [astro-ph]].

\bibitem{cobe}
C.~L.~Bennett {\it et al.},
Astrophys.\ J.\  {\bf 464}, L1 (1996) [arXiv:astro-ph/9601067].

\bibitem{wmap1}
D.~N.~Spergel {\it et al.}  [WMAP Collaboration],
Astrophys.\ J.\ Suppl.\  {\bf 148}, 175 (2003)
[arXiv:astro-ph/0302209].

\bibitem{lowl} 
A.~de Oliveira-Costa, M.~Tegmark, M.~Zaldarriaga and A.~Hamilton,
Phys.\ Rev.\  D {\bf 69}, 063516 (2004) [arXiv:astro-ph/0307282].;
G.~Efstathiou,
Mon.\ Not.\ Roy.\ Astron.\ Soc.\  {\bf 348}, 885 (2004)
[arXiv:astro-ph/0310207];
C.~Copi, D.~Huterer, D.~Schwarz and G.~Starkman,
Phys.\ Rev.\  D {\bf 75}, 023507 (2007) [arXiv:astro-ph/0605135].
%

\bibitem{axis}
K.~Land and J.~Magueijo,
Phys.\ Rev.\ Lett.\  {\bf 95}, 071301 (2005)
[arXiv:astro-ph/0502237].;
%
T.~R.~Jaffe, A.~J.~Banday, H.~K.~Eriksen, K.~M.~Gorski and
F.~K.~Hansen,
Astrophys.\ J.\  {\bf 629}, L1 (2005) [arXiv:astro-ph/0503213].;

\bibitem{cold}
P.~Vielva, E.~Martinez-Gonzalez, R.~B.~Barreiro, J.~L.~Sanz and L.~Cayon,
Astrophys.\ J.\  {\bf 609}, 22 (2004)
[arXiv:astro-ph/0310273].

\bibitem{asym}
F.~K.~Hansen, P.~Cabella, D.~Marinucci and N.~Vittorio,
Astrophys.\ J.\  {\bf 607}, L67 (2004) [arXiv:astro-ph/0402396].;
%
H.~K.~Eriksen, F.~K.~Hansen, A.~J.~Banday, K.~M.~Gorski and
P.~B.~Lilje,
Astrophys.\ J.\  {\bf 605}, 14 (2004) [Erratum-ibid.\  {\bf 609},
1198 (2004)] [arXiv:astro-ph/0307507].;
%
F.~K.~Hansen, A.~J.~Banday and K.~M.~Gorski,
Mon.\ Not.\ Roy.\ Astron.\ Soc.\  {\bf 354}, 641 (2004)
[arXiv:astro-ph/0404206].

\bibitem{Groeneboom:2008fz}
N.~E.~Groeneboom and H.~K.~Eriksen,
Astrophys.\ J.\  {\bf 690}, 1807 (2009)
[arXiv:0807.2242 [astro-ph]].

\bibitem{Gumrukcuoglu:2006xj}
  A.~E.~Gumrukcuoglu, C.~R.~Contaldi and M.~Peloso,
  arXiv:astro-ph/0608405.
Proceeding of the 
``11th Marcel Grossmann Meeting On General Relativity''
Ed. H. Kleinert, R.T. Jantzen and R. Ruffini. Hackensack, World Scientific, 2008;


\bibitem{Ackerman:2007nb}
  L.~Ackerman, S.~M.~Carroll and M.~B.~Wise,
  Phys.\ Rev.\  D {\bf 75}, 083502 (2007)
  [Erratum-ibid.\  D {\bf 80}, 069901 (2009)]
  [arXiv:astro-ph/0701357].

\bibitem{Himmetoglu:2008zp}
  B.~Himmetoglu, C.~R.~Contaldi and M.~Peloso,
  Phys.\ Rev.\ Lett.\  {\bf 102}, 111301 (2009)
  [arXiv:0809.2779 [astro-ph]].
  
\bibitem{Himmetoglu:2008hx}
  B.~Himmetoglu, C.~R.~Contaldi and M.~Peloso,
  Phys.\ Rev.\  D {\bf 79}, 063517 (2009)
  [arXiv:0812.1231 [astro-ph]].

\bibitem{Carroll:2008br}
  S.~M.~Carroll, C.~Y.~Tseng and M.~B.~Wise,
  arXiv:0811.1086 [astro-ph].

\bibitem{Pullen:2007tu}
  A.~R.~Pullen and M.~Kamionkowski,
  Phys.\ Rev.\  D {\bf 76}, 103529 (2007)
  [arXiv:0709.1144 [astro-ph]].

\bibitem{ArmendarizPicon:2008yr}
  C.~Armendariz-Picon and L.~Pekowsky,
  Phys.\ Rev.\ Lett.\  {\bf 102}, 031301 (2009)
  [arXiv:0807.2687 [astro-ph]].

\bibitem{Hanson:2009gu}
  D.~Hanson and A.~Lewis,
  Phys.\ Rev.\  D {\bf 80}, 063004 (2009)
  [arXiv:0908.0963 [astro-ph.CO]].

\bibitem{Groeneboom:2009cb}
  N.~E.~Groeneboom, L.~Ackerman, I.~K.~Wehus and H.~K.~Eriksen,
  arXiv:0911.0150 [astro-ph.CO].

\bibitem{Gumrukcuoglu:2007bx}
  A.~E.~Gumrukcuoglu, C.~R.~Contaldi and M.~Peloso,
  JCAP {\bf 0711}, 005 (2007)
  [arXiv:0707.4179 [astro-ph]].

\bibitem{Pereira:2007yy}
  T.~S.~Pereira, C.~Pitrou and J.~P.~Uzan,
  JCAP {\bf 0709}, 006 (2007)
  [arXiv:0707.0736 [astro-ph]].

\bibitem{Mukhanov:1990me}
  V.~F.~Mukhanov, H.~A.~Feldman and R.~H.~Brandenberger,
  Phys.\ Rept.\  {\bf 215}, 203 (1992).

\bibitem{Gumrukcuoglu:2008gi}
  A.~E.~Gumrukcuoglu, L.~Kofman and M.~Peloso,
  Phys.\ Rev.\  D {\bf 78}, 103525 (2008)
  [arXiv:0807.1335 [astro-ph]].




\bibitem{wald}
R.~W.~Wald,
Phys.\ Rev.\  D {\bf 28}, 2118 (1983).



\bibitem{barrow}
J.~D.~Barrow and S.~Hervik,
Phys.\ Rev.\  D {\bf 73}, 023007 (2006) [arXiv:gr-qc/0511127];
%
Phys.\ Rev.\  D {\bf 74}, 124017 (2006) [arXiv:gr-qc/0610013].
  J.~D.~Barrow and S.~Hervik,
  Phys.\ Rev.\  D {\bf 81}, 023513 (2010)
  [arXiv:0911.3805 [gr-qc]].


\bibitem{nemanja}
N.~Kaloper,
Phys.\ Rev.\  D {\bf 44}, 2380 (1991).

\bibitem{pforms}
E.~Di Grezia, G.~Esposito, A.~Funel, G.~Mangano and G.~Miele,
Phys.\ Rev.\  D {\bf 68}, 105012 (2003)
[arXiv:gr-qc/0305050].
%
  C.~Germani and A.~Kehagias,
  JCAP {\bf 0903}, 028 (2009)
  [arXiv:0902.3667 [astro-ph.CO]].
  T.~Kobayashi and S.~Yokoyama,
  JCAP {\bf 0905}, 004 (2009)
  [arXiv:0903.2769 [astro-ph.CO]].
  T.~S.~Koivisto and N.~J.~Nunes,
  arXiv:0907.3883 [astro-ph.CO].
  C.~Germani and A.~Kehagias,
  JCAP {\bf 0911}, 005 (2009)
  [arXiv:0908.0001 [astro-ph.CO]].
  T.~S.~Koivisto and N.~J.~Nunes,
  Phys.\ Rev.\  D {\bf 80}, 103509 (2009)
  [arXiv:0908.0920 [astro-ph.CO]].
  T.~S.~Koivisto, D.~F.~Mota and C.~Pitrou,
  JHEP {\bf 0909}, 092 (2009)
  [arXiv:0903.4158 [astro-ph.CO]].

\bibitem{ford}
L.~H.~Ford,
Phys.\ Rev.\  D {\bf 40}, 967 (1989).

\bibitem{nongauss}
%
  M.~Karciauskas, K.~Dimopoulos and D.~H.~Lyth,
  Phys.\ Rev.\  D {\bf 80}, 023509 (2009)
  [arXiv:0812.0264 [astro-ph]].
%
  N.~Bartolo, E.~Dimastrogiovanni, S.~Matarrese and A.~Riotto,
  JCAP {\bf 0910}, 015 (2009)
  [arXiv:0906.4944 [astro-ph.CO]].
%
  C.~A.~Valenzuela-Toledo, Y.~Rodriguez and D.~H.~Lyth,
  Phys.\ Rev.\  D {\bf 80}, 103519 (2009)
  [arXiv:0909.4064 [astro-ph.CO]].
%
  N.~Bartolo, E.~Dimastrogiovanni, S.~Matarrese and A.~Riotto,
  JCAP {\bf 0911}, 028 (2009)
  [arXiv:0909.5621 [astro-ph.CO]].
%
  C.~A.~Valenzuela-Toledo and Y.~Rodriguez,
  arXiv:0910.4208 [astro-ph.CO].
%
E.~Dimastrogiovanni, N.~Bartolo, S.~Matarrese, A.~Riotto,
  arXiv:1001.4049 [astro-ph.CO].

\bibitem{Dimopoulos:2006ms}
  K.~Dimopoulos,
  Phys.\ Rev.\  D {\bf 74}, 083502 (2006)
  [arXiv:hep-ph/0607229].

\bibitem{dark}
C.~Armendariz-Picon,
JCAP {\bf 0407}, 007 (2004) [arXiv:astro-ph/0405267].
%
T.~Koivisto and D.~F.~Mota,
Astrophys.\ J.\  {\bf 679}, 1 (2008)
[arXiv:0707.0279 [astro-ph]];
%
T.~S.~Koivisto and D.~F.~Mota,
JCAP {\bf 0808}, 021 (2008) [arXiv:0805.4229 [astro-ph]].
%
J.~B.~Jimenez and A.~L.~Maroto,
Phys.\ Rev.\  D {\bf 80}, 063512 (2009)
[arXiv:0905.1245 [astro-ph.CO]].

\bibitem{Golovnev:2008cf}
  A.~Golovnev, V.~Mukhanov and V.~Vanchurin,
  JCAP {\bf 0806}, 009 (2008)
  [arXiv:0802.2068 [astro-ph]].
  S.~Kanno, M.~Kimura, J.~Soda and S.~Yokoyama,
  JCAP {\bf 0808}, 034 (2008)
  [arXiv:0806.2422 [hep-ph]].
  
  
\bibitem{AR}
  M.~S.~Turner and L.~M.~Widrow,
  Phys.\ Rev.\  D {\bf 37}, 2743 (1988).
  K.~Dimopoulos and M.~Karciauskas,
  JHEP {\bf 0807}, 119 (2008)
  [arXiv:0803.3041 [hep-th]].


\bibitem{Himmetoglu:2009qi}
  B.~Himmetoglu, C.~R.~Contaldi and M.~Peloso,
  Phys.\ Rev.\  D {\bf 80}, 123530 (2009)
  [arXiv:0909.3524 [astro-ph.CO]].

\bibitem{Dimopoulos:2008yv}
  K.~Dimopoulos, M.~Karciauskas, D.~H.~Lyth and Y.~Rodriguez,
  JCAP {\bf 0905}, 013 (2009)
  [arXiv:0809.1055 [astro-ph]].

\bibitem{Golovnev:2009rm}
  A.~Golovnev,
  Phys.\ Rev.\  D {\bf 81}, 023514 (2010)
  [arXiv:0910.0173 [astro-ph.CO]].

\bibitem{stability}
  S.~M.~Carroll, T.~R.~Dulaney, M.~I.~Gresham and H.~Tam,
  Phys.\ Rev.\  D {\bf 79}, 065011 (2009)
  [arXiv:0812.1049 [hep-th]].
  S.~M.~Carroll, T.~R.~Dulaney, M.~I.~Gresham and H.~Tam,
  Phys.\ Rev.\  D {\bf 79}, 065012 (2009)
  [arXiv:0812.1050 [hep-th]].
  G.~Esposito-Farese, C.~Pitrou and J.~P.~Uzan,
  arXiv:0912.0481 [gr-qc].


























\bibitem{Watanabe:2009ct}
  M.~a.~Watanabe, S.~Kanno and J.~Soda,
  Phys.\ Rev.\ Lett.\  {\bf 102}, 191302 (2009)
  [arXiv:0902.2833 [hep-th]].

\bibitem{Kanno:2009ei}
  S.~Kanno, J.~Soda and M.~a.~Watanabe,
  JCAP {\bf 0912}, 009 (2009)
  [arXiv:0908.3509 [astro-ph.CO]].


\bibitem{Demozzi:2009fu}
  V.~Demozzi, V.~Mukhanov and H.~Rubinstein,
  JCAP {\bf 0908}, 025 (2009)
  [arXiv:0907.1030 [astro-ph.CO]].


\bibitem{Dimopoulos:2009am}
  K.~Dimopoulos, M.~Karciauskas and J.~M.~Wagstaff,
  arXiv:0907.1838 [hep-ph].
%
  K.~Dimopoulos, M.~Karciauskas and J.~M.~Wagstaff,
  Phys.\ Lett.\  B {\bf 683}, 298 (2010)
  [arXiv:0909.0475 [hep-ph]].


\bibitem{nonst-kin}
  K.~Bamba, S.~Nojiri and S.~D.~Odintsov,
  Phys.\ Rev.\  D {\bf 77}, 123532 (2008)
  [arXiv:0803.3384 [hep-th]].
%
  S.~Yokoyama and J.~Soda,
  JCAP {\bf 0808}, 005 (2008)
  [arXiv:0805.4265 [astro-ph]].
%
  K.~Bamba and S.~Nojiri,
  arXiv:0811.0150 [hep-th].
%
  S.~Koh and B.~Hu,
  arXiv:0901.0429 [hep-th].
%
  C.~Armendariz-Picon and A.~Diez-Tejedor,
  JCAP {\bf 0912}, 018 (2009)
  [arXiv:0904.0809 [astro-ph.CO]].

\bibitem{Himmetoglu:2009mk}
  B.~Himmetoglu,
  arXiv:0910.3235 [astro-ph.CO].

\bibitem{Contaldi:2003zv}
  C.~R.~Contaldi, M.~Peloso, L.~Kofman and A.~D.~Linde,
  JCAP {\bf 0307}, 002 (2003)
  [arXiv:astro-ph/0303636].
\bibitem{Riotto:2002yw}
  A.~Riotto,
  arXiv:hep-ph/0210162.




\bibitem{Nilles:2001fg}
  H.~P.~Nilles, M.~Peloso and L.~Sorbo,
  JHEP {\bf 0104}, 004 (2001)
  [arXiv:hep-th/0103202].


\bibitem{Dulaney:2010sq}
  T.~R.~Dulaney and M.~I.~Gresham,
  arXiv:1001.2301 [astro-ph.CO].










\end{thebibliography}
\end{document}